\documentclass{jpp}
\usepackage{graphicx}

\usepackage[utf8]{inputenc}
\usepackage[T1]{fontenc}
\usepackage{amsmath}
\usepackage{subfigure}
\usepackage{physics}
\usepackage{bm}
\usepackage[dvipsnames]{xcolor}
\usepackage{upgreek}

\DeclareMathOperator\erfi{erfi}

\DeclareMathOperator{\Gammaf}{\Upgamma}

\shorttitle{Strong gradients in the plateau regime}
\shortauthor{S. Trinczek, F. I. Parra, P. J. Catto, I. Calvo}

\title{Strong gradient neoclassical transport in the plateau regime}

\author{Silvia Trinczek \aff{1}
  \corresp{\email{strincze@pppl.gov}},
  Felix I. Parra \aff{1},
  Peter J. Catto \aff{2}, Iván Calvo \aff{3}
  }

\affiliation{
\aff{1}Princeton Plasma Physics Laboratory, Princeton, NJ 08543, USA
\aff{2}Plasma Science and Fusion Center, Massachusetts Institute of Technology, Cambridge, MA, USA
\aff{3} Laboratorio Nacional de Fusión, CIEMAT, Madrid, 28040, Spain}

\begin{document}

\maketitle

\begin{abstract}
Strong gradient regions in tokamaks such as the pedestal or internal transport barriers are regions of reduced turbulence where neoclassical transport can play a dominant role. 
In pedestals, gradient lengths comparable to the ion poloidal gyroradius have been measured.
Standard neoclassical theory can miss important strong gradient effects in these regions because it assumes that the gradient length scales of density, temperature and potential are larger than the ion poloidal gyroradius. 
We extend plateau regime neoclassical theory into regions of gradients of the order of the ion poloidal gyroradius to capture strong gradient effects on transport processes in the pedestal and internal transport barriers. The fundamental idea behind our new framework is to keep a scale separation between the orbit widths and the gradient length scales by performing a large aspect ratio expansion. In the plateau regime, strong gradients cause poloidal variation that is in-out as well as up-down asymmetric. We study two different test cases assuming either radial force balance or the absence of turbulence and show that strong gradient effects can enhance or reduce standard neoclassical theory predictions in the plateau regime in strong gradient regions. 
\end{abstract}

\section{Introduction}
Neoclassical transport, which is the collisional transport due to toroidicity in tokamaks and stellarators, has been extensively discussed \citep{hinton1976, helander2005}. One of the main assumptions in these studies is that the gradient length scales of density, temperature and potential are much larger than the ion poloidal gyroradius. Neoclassical theory as described by \cite{hinton1976} and \cite{helander2005} is thus limited to regions of weak gradients like the core of a tokamak, where transport is usually dominated by turbulence.\par 

In the pedestal, where gradients are much larger than in the core, ion heat transport has been observed in ASDEX-U to be on the order of neoclassical predictions \citep{Viezzer2018, urano2005}. Similar observations have been made for internal transport barriers in JET \citep{tala2001}. Gradient length scales in transport barriers have been measured to be of the order of the ion poloidal gyroradius \citep{mcdermott2009, Viezzer2013, strait1995}. Thus, the weak gradient assumption of neoclassical theory breaks down precisely where neoclassical transport becomes important.\par

\cite{Trinczek2023} and \cite{trinczek2025a} extended neoclassical theory into strong gradient regions for the case of a large aspect ratio tokamak with collisionality in the banana regime. This new strong gradient neoclassical theory predicts either enhanced or reduced transport, depending on the profiles of density, temperature and mean parallel flow. \par 

Pedestals as observed in Alcator-C Mod and DIII-D tend to be of sufficiently high collisionality to be close to the banana-plateau transition \citep{marr2010, callen2010}. To address these higher collisionality pedestals, we present an extension of neoclassical transport in strong gradient regions for the plateau regime that follows the same approach as \cite{Trinczek2023} and \cite{trinczek2025a}. \par 

\cite{Seol2012} aimed to extend neoclassical theory into regions of strong gradients in the plateau regime, but strong mean parallel flow and mean parallel flow gradients were neglected. \cite{Pusztai2010} and \cite{Catto2011} had the same objective, but assumed weak temperature gradients as well as weak mean parallel flow and zero neoclassical ion particle flux. Both approaches neglected the poloidal variation in the electric potential that forms in this ordering and has a significant effect on the radial fluxes of energy and particles \citep{trinczek2025a, trinczek2026}. We compare our findings with previous work, correct mistakes made in \cite{Pusztai2010} and \cite{Seol2012} and conclude that our model is more comprehensive. \par

In our approach, we keep gradient length scales of the order of the ion poloidal gyroradius for density, potential and temperature and expand in small inverse aspect ratio. We keep the poloidal variation of the electric potential. This poloidal variation modifies transport equations and leads to up-down as well as in-out asymmetry in the plateau regime. The mean parallel flow is allowed to be of the order of the ion thermal velocity. The resulting transport equations explicitly depend on the mean parallel flow. By including sources in the drift kinetic equation, we allow for the possibility of turbulence in the system. Due to these sources, the ion neoclassical particle flux can be much larger than the electron neoclassical particle flux. We derive transport relations for ions and electrons and a formula for the bootstrap current that are modified by strong gradient effects. We also use example profiles to study the modified transport relations and demonstrate that strong gradient effects can enhance or decrease neoclassical transport depending on the input profiles. This is in disagreement with \cite{Seol2012} who claim that strong gradient effects only ever decrease the neoclassical ion energy flux in the plateau regime in comparison to weak gradient theory. \par

We start in section \ref{sec: General equations} with a derivation of neoclassical transport relations for ions and electrons as well as the bootstrap current for collisionality in the boundary between the plateau and the banana regime. We take the collisionality limit for the plateau regime in section \ref{sec: Plateau}. We choose a set of example profiles for densities and temperatures of ions and electrons and solve the transport equations assuming either force balance or neoclassical ambipolarity to determine the radial electric field in section \ref{sec: Case study}. We summarise our results in section \ref{sec: Conclusion}.

\section{General equations} \label{sec: General equations}
We are interested in generalising neoclassical transport theory to allow for gradients of order $L_{n,T,\Phi}\sim\rho_p$, where the gradient length scales are defined as $L_Q\equiv \abs{\partial \ln Q/\partial r}^{-1}$, $n$ is the density, $T$ the ion temperature, $\Phi$ the electric potential, $r$ is the minor radius and $\rho_p$ is the ion poloidal gyroradius.  \par 
We start by deriving equations for the distribution function in the freely passing and trapped--barely passing region before we take moments of the drift kinetic equation to find particle, momentum and energy transport of ions and electrons. Throughout this section, we assume 
\begin{equation}
    \nu_\ast\equiv \frac{qR\nu}{\epsilon^{3/2}v_{t}}  \sim 1
\end{equation} 
and take the subsidiary limit $\nu_\ast\gg1$ for the plateau regime in section \ref{sec: Plateau}. Here, $q$ is the safety factor, $R$ is the major radius, $\nu$ is the ion-ion collision frequency, $\epsilon\equiv r/R$ is the inverse aspect ratio and $v_{t}\equiv \sqrt{2T/m}$ is the ion thermal speed with ion mass $m$. \par 
We work in the limit of small inverse aspect ratio $\epsilon\ll 1$. This implies that the gradient length scales, which we assume to be of the order of the ion poloidal gyroradius, $L_{n,T,\Phi}\sim\rho_p$, are still much larger than the ion Larmor radius $\rho\ll\rho_p\sim q\rho/\epsilon$. In this limit, $\rho_\ast\equiv \rho/L_{n,T,\Phi}\sim\epsilon$. Thus, we can use the drift kinetic equation, given by
\begin{multline}\label{DKE original}
    \left( v_\parallel \bm{\hat{b}}+\bm{v}_E\right)\bcdot \bnabla\theta \pdv{f}{\theta}+\left(\bm{v}_E+\bm{v}_M\right)\bcdot\bnabla\psi\pdv{f}{\psi}\\
    +\left[ \bm{\hat{b}}+\frac{v_\parallel}{\Omega}\bm{\hat{b}}\cross\left(\bm{\hat{b}}\bcdot\bnabla \bm{\hat{b}}\right)\right]\bcdot\left(-\mu\bnabla B+\frac{Ze}{m}\bm{E}\right)\pdv{f}{v_\parallel}=C[f,f]+\Sigma.
\end{multline}
Here, $\bm{v}_E=c\bm{E}\cross\bm{B}/B^2$ is the $\bm{E}\cross\bm{B}$-drift, $\bm{v}_M=\mu \bm{\hat{b}}\cross\bnabla B/\Omega+v_\parallel^2 \bm{\hat{b}}\cross(\bm{\hat{b}}\bcdot \bnabla\bm{\hat{b}})/\Omega$ is the magnetic drift, $C[f,f]$ is the Fokker-Planck ion-ion collision operator and $\Sigma\sim \epsilon^2 v_{t} f/(qR)$ is a source. The electric field $\bm{E}=-\bnabla \Phi$ is electrostatic and the electric potential can be split up into $\Phi(\psi,\theta)=\phi(\psi)+\phi_\theta(\psi,\theta)$, where the poloidally varying piece of the electric potential is small, $\phi_\theta/\phi\sim \epsilon$. The poloidal angle is $\theta$ and $\psi$ is the poloidal flux divided by $2\pi$. The magnetic field is $\bm{B}$, its strength is $B$ and its direction $\bm{\hat{b}}\equiv \bm{B}/B$, $c$ is the speed of light, $\mu=mv_\perp^2/2B$ is the magnetic moment, $v_\perp$ is the perpendicular velocity, $v_\parallel$ is the parallel velocity, $\Omega=ZeB/(mc)$ is the Larmor frequency and $Ze$ is the ion charge.\par

For $L_{n,T,\Phi}\sim\rho_p$ and $\epsilon\ll1$, the lowest order distribution function is Maxwellian, as derived in \cite{Trinczek2023}. The distribution function $f$ can be written as
\begin{equation}
    f=f_M+g,
\end{equation}
where
\begin{equation}\label{Maxwellian}
    f_M(v_\parallel,\mu,\psi)=n(\psi)\left(\frac{m}{2\pi T(\psi)}\right)^{3/2}\exp\bigg\lbrace-\frac{m[v_\parallel-V_\parallel(\psi)]^2}{2T(\psi)}-\frac{m\mu B}{T(\psi)}\bigg\rbrace.
\end{equation}
The mean parallel flow $V_\parallel\sim v_{t}$ is included in the Maxwellian to lowest order and can be of the order of the ion thermal velocity. The piece $g$ will be shown to be small in $\sqrt{\epsilon}$. For a tokamak with concentric circular flux surfaces, we can simplify the drift kinetic equation to lowest order in $\epsilon\sim\rho_\ast\ll1$ and $\rho_p/R\ll 1$ by using
\begin{equation}
    \left( v_\parallel \bm{\hat{b}}+\bm{v}_E\right)\bcdot \bnabla\theta\simeq \frac{v_\parallel+u}{qR},
\end{equation}
\begin{equation}
    \left(\bm{v}_E+\bm{v}_M\right)\bcdot\bnabla\psi\simeq -\frac{I}{\Omega}\frac{v_\parallel^2+\mu B}{q R}\frac{r}{R}\sin\theta -\frac{Ic}{qRB}\pdv{\phi_\theta}{\theta}, 
\end{equation}
and 
\begin{equation}
    \left[ \bm{\hat{b}}+\frac{v_\parallel}{\Omega}\bm{\hat{b}}\cross\left(\bm{\hat{b}}\bcdot\bnabla \bm{\hat{b}}\right)\right]\bcdot\left(-\mu\bnabla B+\frac{Ze}{m}\bm{E}\right)\simeq\frac{v_\parallel u-\mu B}{qR}\frac{r}{R}\sin\theta-\frac{Ze}{qRm}\pdv{\phi_\theta}{\theta}.
\end{equation}
Here, $I\equiv RB_\zeta$, $B_\zeta$ is the toroidal component of the magnetic field, and we used that $B\simeq B_0\left[1-(r/R)\cos\theta\right]$ for concentric circular flux surfaces, where $B_0$ is the magnetic field strength on the magnetic axis. 
We also introduced the velocity
\begin{equation}\label{def u}
    u(\psi,\theta)\equiv \frac{cI}{B}\pdv{\Phi}{\psi}=\frac{cI}{B}\pdv{\phi(\psi)}{\psi}+\frac{cI}{B}\pdv{\phi_\theta(\psi,\theta)}{\psi},
\end{equation}
where $u\sim v_{t}$. This is the parallel speed of the trapped particles and it is connected to the poloidal projection of the $E\times B$-drift. Note that this definition is slightly different from that in \cite{Trinczek2023}, where $u$ was defined to be proportional to the derivative of only the poloidally independent part of the electric potential. The difference between the two definitions will turn out to be negligible for most of our discussion. With these results and recalling that $\rho_\ast\sim\epsilon\ll1$ and $\nu_\ast\sim1$, the drift kinetic equation \eqref{DKE original} for ions becomes
\begin{multline}\label{drift g}
    \frac{v_\parallel+u}{qR}\pdv{g}{\theta}+\left(\frac{v_\parallel u-\mu B}{qR}\frac{r}{R}\sin\theta-\frac{Ze}{qRm}\pdv{\phi_\theta}{\theta}\right)\pdv{g}{v_\parallel}\\
    -\frac{I}{\Omega}\left(\frac{v_\parallel^2+\mu B}{q R}\frac{r}{R}\sin \theta+\frac{Ze}{qRm}\pdv{\phi_\theta}{\theta}\right)\pdv{g}{\psi}\\
    =\frac{I}{\Omega}\left(\frac{v_\parallel^2+\mu B}{q R}\frac{r}{R}\sin\theta+\frac{Ze}{qRm}\pdv{\phi_\theta}{\theta}\right)\Bigg[\pdv{}{\psi}\ln p+\left(\frac{m(v_\parallel-V_\parallel)^2}{2T}+\frac{m\mu B}{T}-\frac{5}{2}\right)\pdv{}{\psi}\ln T\\
    +\frac{m(v_\parallel-V_\parallel)}{T}\left(\pdv{V_\parallel}{\psi}-\frac{\Omega}{I}\right)\Bigg]f_M+\frac{v_\parallel+u}{qR}\frac{r}{R}\sin\theta\frac{m}{T}\left[v_\parallel(v_\parallel-V_\parallel)+\mu B\right]f_M+C^{(l)}[g]+\Sigma,
\end{multline}
with the ion pressure $p=nT$. The linearised collision operator for ions is
\begin{equation} \label{collision operator}
\begin{split}
   C^{(l)}[g] = & \lambda \bnabla_v\bcdot\left[\int\text{d}^3v'f_M f'_M \bnabla_\omega\bnabla_\omega \omega \bcdot\left(\bnabla_v\left(\frac{g}{f_M}\right)-\bnabla_{v'}\left(\frac{g'}{f_M'}\right)\right)\right]\\
    \simeq & \bnabla_v\bcdot\Bigg[f_M \mathsfbi{M} \bcdot\bnabla_v\left(\frac{g}{f_M}\right)-\lambda f_M\int_{V_{tbp}}\mathrm{d}^3v' f'_M \bnabla_\omega\bnabla_\omega \omega \bcdot \bnabla_v'\left(\frac{g^{t'}}{f'_M}\right)\\
    &-\lambda f_M\int_{V_{p}}\mathrm{d}^3v' f'_M \bnabla_\omega\bnabla_\omega \omega \bcdot \bnabla_v'\left(\frac{g^{p'}}{f'_M}\right)\Bigg]\\
    &\equiv \bnabla_v\bcdot\bm{M},
\end{split}
\end{equation}
where the subscript $V_{tbp}$ indicates that the integral is over the trapped--barely passing region of velocity space ($\abs{v_\parallel+u}\sim\sqrt{\epsilon}v_t$) and similarly the subscript $V_p$ means that the integral is over the freely passing region ($\abs{v_\parallel+u}\gg\sqrt{\epsilon}v_t$), $\lambda=2\pi Z^4e^4 \log\Lambda$, $\boldsymbol{\omega}=\boldsymbol{v}-\boldsymbol{v}'$, $\omega=\abs{\bm{\omega}}$ and $\log\Lambda$ is the Coulomb logarithm. We introduce the matrix
\begin{equation}
    \mathsfbi{M}=\frac{\nu_\perp}{4}\left(|\boldsymbol{v}-V_\parallel\boldsymbol{\hat{b}}|^2\mathsfbi{I}-(\boldsymbol{v}-V_\parallel\boldsymbol{\hat{b}})(\boldsymbol{v}-V_\parallel\boldsymbol{\hat{b}})\right)+\frac{\nu_\parallel}{2}(\boldsymbol{v}-V_\parallel\boldsymbol{\hat{b}})(\boldsymbol{v}-V_\parallel\boldsymbol{\hat{b}}),
\end{equation}
\begin{align}
    \nu_\perp=3\sqrt{\frac{\pi}{2}}\nu\frac{\Xi(x)-\Psi(x)}{x^3}, &&
    \nu_\parallel=3\sqrt{\frac{\pi}{2}}\nu\frac{\Psi(x)}{x^3},&&\text{and}&&
    \nu= \frac{4\sqrt{\pi} Z^4e^4n\log\Lambda}{3T^{3/2}m^{1/2}},
\end{align}
where $x=\sqrt{m/(2T)}|\boldsymbol{v}-V_\parallel\boldsymbol{\hat{b}}|$, $\Xi(x)=\text{erf}(x)$ and $\Psi(x)=(\Xi - x\Xi')/(2x^2)$ is the Chandrasekhar function. 
\par
For some calculations, it is convenient to write the left hand side of the drift kinetic equation in \eqref{drift g} in a conservative form,
\begin{multline}\label{drift g conservative}
  \pdv{}{\theta}\left(\frac{v_\parallel+u}{qR}g\right)+\pdv{}{v_\parallel}\left[\left(\frac{v_\parallel u-\mu B}{qR}\frac{r}{R}\sin\theta-\frac{Ze}{qRm}\pdv{\phi_\theta}{\theta}\right)g\right]\\
    -\pdv{}{\psi}\left[\frac{I}{\Omega}\left(\frac{v_\parallel^2+\mu B}{q R}\frac{r}{R}\sin \theta+\frac{Ze}{qRm}\pdv{\phi_\theta}{\theta}\right)g\right]\\
    =\frac{I}{\Omega}\left(\frac{v_\parallel^2+\mu B}{q R}\frac{r}{R}\sin\theta+\frac{Ze}{qRm}\pdv{\phi_\theta}{\theta}\right)\Bigg[\pdv{}{\psi}\ln p+\left(\frac{m(v_\parallel-V_\parallel)^2}{2T}+\frac{m\mu B}{T}-\frac{5}{2}\right)\pdv{}{\psi}\ln T\\
    +\frac{m(v_\parallel-V_\parallel)}{T}\left(\pdv{V_\parallel}{\psi}-\frac{\Omega}{I}\right)\Bigg]f_M+\frac{v_\parallel+u}{qR}\frac{r}{R}\sin\theta\frac{m}{T}\left[v_\parallel(v_\parallel-V_\parallel)+\mu B\right]f_M+C^{(l)}[g]+\Sigma.
\end{multline}
Note that to obtain this expression, we have used $\partial u/\partial\theta\simeq -u(r/R)\sin\theta+(cI/B)(\partial^2\phi_\theta/\partial\psi\partial\theta)$. For this relation to hold, we needed to keep $\phi_\theta$ in the definition of $u$ in \eqref{def u}.
\par As discussed in detail in \cite{Trinczek2023}, \cite{trinczek2025a} and \cite{trinczek2026}, trapped particles are particles with velocity $v_\parallel\simeq -u$. For this reason, it is useful to make a change of variables from $v_\parallel$ to $w\equiv v_\parallel+u$ in the trapped--barely passing region,
\begin{equation}
 \pdv{g}{\psi}\Bigg\vert_{\theta,v_\parallel}=\pdv{g}{\psi}\Bigg\vert_{\theta,w}+\pdv{u}{\psi}\Bigg\vert_{\theta}\pdv{g}{w}\Bigg\rvert_{\theta,\psi},
 \end{equation}
\begin{equation}
 \pdv{g}{v_\parallel}\Bigg\rvert_{\theta,\psi}=\pdv{g}{w}\Bigg\rvert_{\theta,\psi},
\end{equation}
\begin{equation}\label{dtheta change}
    \pdv{g}{\theta}\Bigg\rvert_{v_\parallel,\psi}=\pdv{g}{\theta}\Bigg\rvert_{w,\psi}+\pdv{u}{\theta}\Bigg\rvert_{\psi}\pdv{g}{w}\Bigg\rvert_{\theta,\psi}.
\end{equation}
The change of variables from $v_\parallel$ to $w$ is convenient to distinguish the trapped--barely passing and the freely passing region. In the freely passing region, $w\sim v_{t}$ and hence we assume that $\partial g^p/\partial w\sim g^p/v_{t}$, whereas in the trapped-barely passing region $w\sim \epsilon^{1/2} v_{t}$ and thus $\partial g^{t,bp}/\partial w \sim g^{t,bp}/(\epsilon^{1/2} v_{t})$. Here, $g^p$ denotes the distribution function of freely passing particles and $g^{t,bp}$ is the distribution function in the trapped and barely passing region, respectively. Note that the second term in \eqref{dtheta change} is small in $\epsilon$ because $u$ depends on $\theta$ through the derivatives of $\phi_\theta$ and $B$, as can be seen from \eqref{def u}. In the trapped--barely passing region, we will use the conservative form of the drift kinetic equation in the variables $(\psi,\theta, w,\mu)$,
\begin{multline}\label{drift g conservative trapped}
  \pdv{}{\theta}\left(\frac{w}{qR}g\right)+\pdv{}{w}\left(\pdv{u}{\theta}\frac{1}{qR}w g\right)    -\pdv{}{\psi}\left[\frac{I}{\Omega}\left(\frac{(w-u)^2+\mu B}{q R}\frac{r}{R}\sin \theta+\frac{Ze}{qRm}\pdv{\phi_\theta}{\theta}\right)g\right]\\
   +\pdv{}{w}\left[S\left(\frac{(w-u) u-\mu B}{qR}\frac{r}{R}\sin\theta-\frac{Ze}{qRm}\pdv{\phi_\theta}{\theta}\right)g-\frac{I}{\Omega}\pdv{u}{\psi}\frac{(w-u) w}{qR}\frac{r}{R}\sin\theta g\right]\\
    -\frac{I}{\Omega}\left(\frac{(w-u)^2+\mu B}{q R}\frac{r}{R}\sin\theta+\frac{Ze}{qRm}\pdv{\phi_\theta}{\theta}\right)\Bigg[\pdv{}{\psi}\ln p+\frac{m(w-u-V_\parallel)}{T}\left(\pdv{V_\parallel}{\psi}-\frac{\Omega}{I}\right)\\
    +\left(\frac{m(w-u-V_\parallel)^2}{2T}+\frac{m\mu B}{T}-\frac{5}{2}\right)\pdv{}{\psi}\ln T
    \Bigg]f_M\\
    -\frac{w}{qR}\frac{r}{R}\sin\theta\frac{m}{T}\left[(w-u)(w-u-V_\parallel)+\mu B\right]f_M=C^{(l)}[g]+\Sigma.
\end{multline}
Here, we introduced the squeezing factor $S$ as
\begin{equation}
    S\equiv1+\frac{cI^2}{B\Omega}\pdv[2]{\Phi}{\psi}.
\end{equation}
\par Going forward, when convenient, we will abbreviate the drift kinetic operator applied on $g$ and $f_M$ as $\mathcal{L}$, allowing us to rewrite \eqref{drift g}, \eqref{drift g conservative} and \eqref{drift g conservative trapped} as
\begin{equation}\label{DK in L}
    \mathcal{L}[g]+\mathcal{L}[f_M]=C^{(l)}[g]+\Sigma.
\end{equation}

\subsection{Distribution function}
We expand the distribution function $g$ in the size of the trapped region $\epsilon^{1/2}$,
\begin{align}
    g=g_0+g_1+...,
\end{align}
where $g_1\sim\epsilon^{1/2}g_0$ and $g_0\sim \epsilon^{1/2}f_M$. This expansion is useful both in the trapped--barely passing region for $g^{t,bp}$ as well as in the freely passing region for $g^p$. The distribution functions have to match asymptotically, i.e.
\begin{equation}\label{matching condition}
    g^p(v_\parallel\rightarrow-u^+)=g^{bp}(w\rightarrow\infty)\quad\text{and}\quad g^p(v_\parallel\rightarrow-u^-)=g^{bp}(w\rightarrow-\infty).
\end{equation}
The distribution function $g^p$ is discontinuous across the trapped--barely passing region. This jump $\Delta g^p$ matches the limits of the barely passing distribution function
\begin{equation}\label{Delta gp}
    \Delta g^p\equiv g^p(v_\parallel\rightarrow u^+)-g^p(v_\parallel\rightarrow u^-)=g^{bp}(w\rightarrow\infty)-g^{bp}(w\rightarrow -\infty).
\end{equation}
\par 

In the freely passing region, $g^p$ is independent of $\theta$ to lowest order in $\sqrt{\epsilon}$, as shown by \cite{Trinczek2023}. The $\theta$-dependent part was calculated by \cite{Trinczek2023} and yields
\begin{equation} \label{theta dependent gp}
    \begin{split}
        &g^p-\langle g^p\rangle_\tau        =-\frac{I}{\Omega}\frac{r}{R}\frac{\left(v_\parallel^2+\mu B\right) \cos{\theta}-ZeR\phi_\theta/mr}{v_\parallel +u}\Bigg[\pdv{}{\psi}\ln{p} +\frac{m(v_\parallel- V_\parallel)}{T}\left(\pdv{V_\parallel}{\psi}-\frac{\Omega}{I}\right)\\
        +&\left(\frac{m(v_\parallel-V_\parallel)^2}{2T}+\frac{m\mu B}{T}-\frac{5}{2}\right)\pdv{}{\psi}\ln{T}\Bigg]f_M-\frac{r}{R}\cos{\theta}\frac{m}{T}\left[v_\parallel(v_\parallel-V_\parallel)+\mu B \right]f_M\sim \epsilon f_M.
    \end{split}
\end{equation}
This expression can also be found by noting that the terms proportional to $\partial g/\partial v_\parallel$ and $\partial g/\partial \psi$ in \eqref{drift g} are negligible in the freely passing region, where $\partial/\partial v_\parallel\sim1/v_t$. The $\theta$-dependence in the trapped--barely passing region has to match the freely passing solution according to \eqref{matching condition}. Thus, any $\theta$-dependence in $g_0^{bp}$ has to decay as $w\rightarrow\pm\infty$ and cannot contribute to the jump $\Delta g^p$.\par

In the trapped--barely passing region, we can use that $w\simeq 0$ to lowest order. The linearised collision operator \eqref{collision operator} to lowest order in $\epsilon$ is
\begin{equation}\label{C lowest order}
    C^{(l)}[g]=\mathsf{M}_\parallel \pdv[2]{g^{t,bp}_0}{w},
\end{equation}
where we defined
\begin{equation}
    \mathsf{M}_\parallel\equiv \bm{\hat{b}}\bcdot\mathsfbi{M}\bcdot\bm{\hat{b}}\Big\rvert_{w=0}\simeq\frac{\nu_\perp}{2}\mu B+\frac{\nu_\parallel}{2}(u+V_\parallel)^2.
\end{equation}
Here, $\nu_\perp$ and $\nu_\parallel$ are evaluated at $w=0$. We find that \eqref{drift g conservative trapped} is, to lowest order in $\sqrt{\epsilon}$ and for $\nu_\ast\sim1$, 
\begin{multline}\label{General Jump Condition}
    \pdv{}{\theta}\left(\frac{w}{qR}g_0^{t,bp}\right)-S\pdv{}{w}\left[\mathcal{P}(\theta)\frac{r}{R}\frac{1}{qR}g_0^{t,bp}\right]\\
    =\frac{I}{\Omega}\mathcal{P}(\theta)\frac{r}{R}\frac{1}{qR}\mathcal{D}f_M(w=0)+\mathsf{M}_\parallel \pdv[2]{g^{t,bp}_0}{w},
\end{multline}
where we introduced the poloidal dependence
\begin{equation}
    \mathcal{P}(\theta)=(u^2+\mu B)\sin\theta+\frac{ZeR}{mr}\pdv{\phi_\theta}{\theta}
\end{equation}
and the derivative term
\begin{equation}
    \mathcal{D}\equiv\pdv{}{\psi}\ln{p} -\frac{m(u+ V_\parallel)}{T}\left(\pdv{V_\parallel}{\psi}-\frac{\Omega}{I}\right) +\left(\frac{m(u+V_\parallel)^2}{2T}+\frac{m\mu B}{T}-\frac{5}{2}\right)\pdv{}{\psi}\ln{T}.
\end{equation}
In the large $w$ limit, \eqref{General Jump Condition} reduces to
\begin{equation}\label{big w limit}
    \pdv{}{\theta}\left(\frac{w}{qR}g_0^{t,bp}\right)
    =\frac{I}{\Omega}\mathcal{P}(\theta)\frac{r}{R}\frac{1}{qR}\mathcal{D}f_M(w=0).
\end{equation}
From this, we find that the $\theta$-dependent piece of $g^{t,bp}_0$ decays as $\sim1/w$ for $w\rightarrow \infty$ as expected. Furthermore, we recover the terms in expression \eqref{theta dependent gp} that diverge as $1/w$ for small $w$. The other terms in \eqref{theta dependent gp} match the large $w$ limit of the next order correction $g_1^{t,bp}$. Equation \eqref{General Jump Condition} is sufficient to determine neoclassical transport quantities and we will not solve for $g_1^{t,bp}$ in this article.

\subsection{Moment equations}
We are interested in calculating particle, parallel momentum and energy transport in the limit of $\epsilon\ll1$ and $\nu_\ast\sim 1$. The transport relations are associated with moments of the drift kinetic equation. The integration has to treat freely passing and trapped--barely passing particles separately. First, we discuss the freely passing region.\par

The transit average of \eqref{drift g} appears to have many contributions by terms proportional to $f_M$ and terms proportional to derivatives of $g$. However, the $\theta$-dependence of $f_M$ and $g^p$ is of order \textit{O}$(\epsilon)$ -- see \eqref{theta dependent gp}. Thus, the transit average of \eqref{drift g} in the freely passing region to lowest order in $\epsilon$ reduces to
\begin{equation} \label{fixed theta DKE}
    \langle C_p^{(l)}[g]\rangle_\tau\simeq-\langle\Sigma\rangle_\tau,
\end{equation}
where $C_p^{(l)}[g]$ is the linearised collision operator in the freely passing region
\begin{multline}
    C_p^{l}[g]= \bnabla_v\bcdot\Bigg[f_M \mathsfbi{M} \bcdot\bnabla_v\left(\frac{g^p}{f_M}\right)-\lambda f_M\int_{V_{tbp}}\mathrm{d}^3v' f'_M \bnabla_\omega\bnabla_\omega \omega \bcdot \bnabla_v'\left(\frac{g^{t,bp'}}{f'_M}\right)\\
    -\lambda f_M\int_{V_{p}}\mathrm{d}^3v' f'_M \bnabla_\omega\bnabla_\omega \omega \bcdot \bnabla_v'\left(\frac{g^{p'}}{f'_M}\right)\Bigg] \equiv \bnabla_v\bcdot\bm{M}^p.
\end{multline}
We define the transit average for passing particles as
\begin{align}
    \langle ...\rangle_\tau=\frac{1}{\tau} \int_0^{2\pi}\mathrm{d}\theta\:\frac{ qR}{v_\parallel+u}(...), 
    \end{align}
    where
    \begin{align}
    \tau=\int_0^{2\pi}\mathrm{d}\theta\:\frac{qR}{v_\parallel+u}.
\end{align}
\cite{Trinczek2023} introduced a new set of variables that is based on conserved quantities, which proved \eqref{fixed theta DKE} to be exact if the new variables are held fixed in the transit average. The difference between the new variables and $\lbrace v_\parallel,\psi\rbrace$ is small in $\epsilon$ for freely passing particles and negligible for the purpose of this paper. We derive transport equations of particles, momentum and energy by taking moments of the drift kinetic equation \eqref{fixed theta DKE}. \par

The integration of \eqref{fixed theta DKE} over the freely passing region to find the particle transport is
\begin{equation}\label{Vp int DKE}
    \int_{V_p}\mathrm{d}^3v\: \langle C_p^{(l)}[g]\rangle_\tau=-\int_{V_p}\mathrm{d}^3v\:\langle\Sigma\rangle_\tau.
\end{equation}
The integration over the freely passing region is defined as
\begin{equation}
    \int_{V_p}\mathrm{d}^3v\:(...)=\int_{-\infty}^{-u^-}\mathrm{d}v_\parallel\:\int_0^\infty\mathrm{d}\mu\:2\pi B(...)+\int_{-u^+}^{\infty}\mathrm{d}v_\parallel\:\int_0^\infty\mathrm{d}\mu\:2\pi B(...),
\end{equation}
where the limits $-u^-$ and $-u^+$ indicate that $g^p$ is not continuous at $-u$ and we need to distinguish between the two sides of $-u$.

The transit average for the freely passing particles can be written as a flux surface average
\begin{equation}
    \langle...\rangle_\psi=\int\frac{\mathrm{d}\theta}{2\pi}(...).
\end{equation}
The integration over $\mu$ in \eqref{Vp int DKE} cancels out the $\mu$ derivative in the divergence of the collision operator and \eqref{Vp int DKE} becomes
\begin{equation}
    \bigg\langle\int_{V_p}\mathrm{d}^3v\:\pdv{}{v_\parallel}\left[ \bm{\hat{b}}\bcdot\bm{M}^p\right]\bigg\rangle_\psi
    \simeq-\bigg\langle\int_{V_p} \mathrm{d}^3 v\:\Sigma\bigg\rangle_\psi.
\end{equation}
The integration over $v_\parallel$ gives a jump contribution,
\begin{equation}\label{passing particle}
      \bigg\langle\int\mathrm{d}\mu\:2\pi B\Delta\left[\bm{\hat{b}}\bcdot\bm{M}^p \right] \bigg\rangle_\psi \simeq\bigg\langle\int_{V_p} \mathrm{d}^3 v\:\Sigma\bigg\rangle_\psi.
\end{equation}
In the freely passing region, the jump of a function $\Delta\mathcal{F}^p$ across the trapped--barely passing region is defined as
\begin{equation}
    \Delta \mathcal{F}^p\equiv \mathcal{F}^p(v_\parallel\rightarrow-u^+)-\mathcal{F}^p(v_\parallel\rightarrow-u^-).
\end{equation}
By integrating the drift kinetic equation over the freely passing region, we connected the particle transport and the associated particle flux to a jump across the trapped--barely passing region. \par

Similarly, we can take the parallel momentum moment of \eqref{fixed theta DKE}
\begin{equation}\label{Vp vpar int DKE}
    \int_{V_p}\mathrm{d}^3v\: mv_\parallel \langle C_p^{(l)}[g]\rangle_\tau=-\int_{V_p}\mathrm{d}^3v\:mv_\parallel \langle\Sigma\rangle_\tau
\end{equation}
to find
\begin{equation}\label{passing momentum}
     \bigg\langle\int\mathrm{d}\mu\:2\pi B\Delta\left[mv_\parallel \bm{\hat{b}}\bcdot\bm{M}^p\right] \bigg \rangle_\psi      +\bigg\langle \int_{V_p}\mathrm{d}^3v\: m\bm{\hat{b}}\bcdot\bm{M}^p\bigg\rangle_\psi\simeq\bigg\langle \int_{V_p} \mathrm{d}^3 v\:mv_\parallel\Sigma\bigg\rangle_\psi.
\end{equation}

\par For the energy moment of \eqref{fixed theta DKE},
\begin{equation}\label{Vp v^2 int DKE}
    \int_{V_p}\mathrm{d}^3v\: \frac{mv^2}{2} \langle C_p^{(l)}[g]\rangle_\tau=-\int_{V_p}\mathrm{d}^3v\:\frac{mv^2}{2} \langle\Sigma\rangle_\tau,
\end{equation}
we find
\begin{equation}\label{passing energy}
     \bigg\langle\int\mathrm{d}\mu\:2\pi B\Delta\left[ \frac{mv^2}{2}\bm{\hat{b}}\bcdot\bm{M}^p\right] \bigg \rangle_\psi      +\bigg\langle \int_{V_p}\mathrm{d}^3v\:m\bm{v}\bcdot  \bm{M}^p\bigg\rangle_\psi\simeq\bigg\langle \int_{V_p} \mathrm{d}^3 v\:\frac{mv^2}{2}\Sigma\bigg\rangle_\psi.
\end{equation}

\par Particle, parallel momentum and energy conservation equations \eqref{passing particle}, \eqref{passing momentum} and \eqref{passing energy} all depend on jump contributions.
We need to determine the jumps from the trapped--barely passing region.\par

For the particle transport, we can integrate the drift kinetic equation in \eqref{drift g conservative trapped} over velocity space and flux surface average. The integration over the collision operator gives the jump contribution,
\begin{multline}\label{trapped jump}
    \bigg\langle \int\mathrm{d}\mu\:2\pi B\Delta\left[\bm{\hat{b}}\bcdot\bm{M}^{t,bp} \right]\bigg\rangle_\psi=-\bigg\langle\int_{V_{tbp}} \mathrm{d}^3v\:\Sigma\bigg\rangle_\psi\\
    + \bigg\langle \int_{V_{tbp}} \mathrm{d}^3v\: \mathcal{L}[g^{t,bp}]\bigg\rangle_\psi+\bigg\langle \int_{V_{tbp}} \mathrm{d}^3v\: \mathcal{L}[f_M]\bigg\rangle_\psi.
\end{multline}
The jump across the trapped--barely passing region in said region is defined as
\begin{equation}
    \Delta \mathcal{F}^{bp}=\mathcal{F}^{bp}(w\rightarrow\infty)-\mathcal{F}^{bp}(w\rightarrow-\infty).
\end{equation}
The jump of a quantity $\mathcal{F}$ has to be the same in the region of overlap between the regions of freely passing and trapped--barely passing particles,
\begin{equation}
    \Delta \mathcal{F}^{bp}=\Delta \mathcal{F}^p,
\end{equation}
and thus
\begin{equation}
     \bigg\langle \int\mathrm{d}\mu\:2\pi B\Delta\left[\bm{\hat{b}}\bcdot\bm{M}^{t,bp} \right]\bigg\rangle_\psi= \bigg\langle \int\mathrm{d}\mu\:2\pi B\Delta\left[\bm{\hat{b}}\bcdot\bm{M}^p \right]\bigg\rangle_\psi.
\end{equation}
The jump \eqref{trapped jump} can be substituted into the particle transport equation \eqref{passing particle} to find
\begin{equation}\label{Particle total}
   \bigg\langle\int \mathrm{d}^3v\:\Sigma\bigg\rangle_\psi = \bigg\langle \int_{V_{tbp}} \mathrm{d}^3v\: \mathcal{L}[g^{t,bp}]\bigg\rangle_\psi+\bigg\langle \int_{V_{tbp}} \mathrm{d}^3v\: \mathcal{L}[f_M]\bigg\rangle_\psi.
\end{equation}
\par
The jump in parallel momentum follows from multiplying \eqref{drift g conservative trapped} by $mv_\parallel$ and integrating over the trapped--barely passing particle velocity space. We find
\begin{multline}\label{trapped momentum jump}
    \bigg\langle \int\mathrm{d}\mu\:2\pi B\Delta\left[mv_\parallel\bm{\hat{b}}\bcdot\bm{M}^{t,bp} \right]\bigg\rangle_\psi=\bigg\langle \int_{V_{tbp}}\mathrm{d}^3v\: m\bm{\hat{b}}\bcdot\bm{M}^{t,bp}\bigg\rangle_\psi-\bigg\langle\int_{V_{tbp}} \mathrm{d}^3v\:mv_\parallel \Sigma\bigg\rangle_\psi\\
    + \bigg\langle \int_{V_{tbp}} \mathrm{d}^3v\: m(w-u)\mathcal{L}[g^{t,bp}]\bigg\rangle_\psi+\bigg\langle \int_{V_{tbp}} \mathrm{d}^3v\: m(w-u)\mathcal{L}[f_M]\bigg\rangle_\psi,
\end{multline}
where we have used that $v_\parallel=w-u$. It is important at this point to keep the distinction between $v_\parallel$ and $-u$ in the trapped--barely passing region. The parallel momentum equation \eqref{passing momentum} with the jump \eqref{trapped momentum jump} becomes

\begin{multline}\label{Momentum total}
   \bigg\langle\int \mathrm{d}^3v\:mv_\parallel \Sigma\bigg\rangle_\psi = \\
   \bigg\langle \int_{V_{tbp}} \mathrm{d}^3v\: m(w-u)\mathcal{L}[g^{t,bp}]\bigg\rangle_\psi+\bigg\langle \int_{V_{tbp}} \mathrm{d}^3v\: m(w-u)\mathcal{L}[f_M]\bigg\rangle_\psi.
\end{multline}
Here, we used that the collision operator conserves momentum,
\begin{equation}\label{Col momentum}
    \int\mathrm{d}^3v\: v_\parallel \bnabla_v\bcdot\bm{M}=-\int\mathrm{d}^3v\: \bm{\hat{b}}\bcdot\bm{M}=-\int_{V_{p}}\mathrm{d}^3v\: \bm{\hat{b}}\bcdot\bm{M}^p-\int_{V_{tbp}}\mathrm{d}^3v\: \bm{\hat{b}}\bcdot\bm{M}^{t,bp}=0.
\end{equation}

\par To find the jump in the energy, we can multiply the drift kinetic equation in \eqref{drift g conservative trapped} by $mv^2/2$ and integrate over the trapped--barely passing velocity space, 
\begin{multline}\label{trapped energy jump}
    \bigg\langle \int\mathrm{d}\mu\:2\pi B\Delta\left[\frac{mv^2}{2}\bm{\hat{b}}\bcdot\bm{M}^{t,bp} \right]\bigg\rangle_\psi=-\bigg\langle \int_{V_{tbp}}\mathrm{d}^3v\:m\bm{v}\bcdot\bm{M}^{t,bp} \bigg\rangle_\psi-\bigg\langle\int_{V_{tbp}} \mathrm{d}^3v\:\frac{mv^2}{2}\Sigma\bigg\rangle_\psi\\
    + \bigg\langle \int_{V_{tbp}} \mathrm{d}^3v\: \frac{mv^2}{2}\mathcal{L}[g^{t,bp}]\bigg\rangle_\psi+\bigg\langle \int_{V_{tbp}} \mathrm{d}^3v\: \frac{mv^2}{2}\mathcal{L}[f_M]\bigg\rangle_\psi.
\end{multline}
The energy moment \eqref{passing energy} gives
\begin{multline}\label{Energy total}
   \bigg\langle\int \mathrm{d}^3v\:\frac{mv^2}{2}\Sigma\bigg\rangle_\psi =   \bigg\langle \int_{V_{tbp}} \mathrm{d}^3v\: \left(\frac{m(w-u)^2}{2}+m\mu B\right)\mathcal{L}[g^{t,bp}]\bigg\rangle_\psi\\+\bigg\langle \int_{V_{tbp}} \mathrm{d}^3v\:\left(\frac{m(w-u)^2}{2}+m\mu B\right) \mathcal{L}[f_M]\bigg\rangle_\psi.
\end{multline}
Here, we used the energy conservation property of the collision operator,
\begin{multline}
     \bigg\langle \int
   \mathrm{d}^3v\:\frac{mv^2}{2}\bnabla_v\bcdot  \bm{M}\bigg\rangle_\psi=-\bigg\langle \int
   \mathrm{d}^3v\:m\bm{v}\bcdot  \bm{M}\bigg\rangle_\psi\\
   =-\bigg\langle \int_{V_{p}}
   \mathrm{d}^3v\:m\bm{v}\bcdot  \bm{M}^p\bigg\rangle_\psi-\bigg\langle \int_{V_{tbp}}
   \mathrm{d}^3v\:m\bm{v}\bcdot  \bm{M}^{t,bp}\bigg\rangle_\psi=0.
\end{multline}

\subsection{Particle, momentum and energy transport}
The particle transport expression \eqref{Particle total} requires evaluating $\langle\int_{V_{tbp}}\mathcal{L}[g^{t,bp}]\mathrm{d}^3v\rangle_\psi$ and $\langle\int_{V_{tbp}}\mathcal{L}[f_M]\mathrm{d}^3v\rangle_\psi$. The integral $\langle\int_{V_{tbp}}\mathcal{L}[f_M]\mathrm{d}^3v\rangle_\psi$ vanishes due to the integral over $\theta$. To lowest order, using expression \eqref{General Jump Condition}, the integral $\langle\int_{V_{tbp}}\mathcal{L}[g^{t,bp}]\mathrm{d}^3v\rangle_\psi$ gives
\begin{equation}\label{lowest Gamma}
     \bigg\langle\int_{V_{tbp}} \mathrm{d}^3v\:\mathcal{L}[g^{t,bp}]\bigg\rangle_\psi \simeq -S\bigg\langle \int_{V_{tbp}} \mathrm{d}^3v\:\frac{1}{qR}\frac{r}{R}\pdv{}{w} \left[\mathcal{P}(\theta)g^{t,bp}_{0}\right]\bigg\rangle_\psi.
\end{equation}
The integral over the $\theta$-independent piece of $g^{t,bp}_0$ vanishes when averaged over the flux surface. The $\theta$-dependent piece of $g^{t,bp}_0$ decays for $w\rightarrow\pm\infty$ in order to match the $\theta$-dependence in the freely passing region as argued below \eqref{theta dependent gp} and shown in \eqref{big w limit}. Thus, the integration over the $\theta$-dependent piece of $g^{t,bp}_0$ vanishes, too. \par
To next order in $\sqrt{\epsilon}$ of \eqref{Particle total}, using expression \eqref{drift g conservative trapped}, we find
\begin{multline}\label{Gamma g1}
    \bigg\langle\int_{V_{tbp}} \mathrm{d}^3v\:\mathcal{L}[g^{t,bp}]\bigg\rangle_\psi \simeq -S\bigg\langle \int_{V_{tbp}} \mathrm{d}^3v\:\frac{1}{qR}\frac{r}{R}\pdv{}{w} \left[\mathcal{P}(\theta)g^{t,bp}_{1}\right]\bigg\rangle_\psi \\
    + \bigg\langle \int_{V_{tbp}} \mathrm{d}^3v\: \pdv{u}{\theta}\frac{1}{qR}\pdv{}{w}\left(w g^{t,bp}_{0}\right) \bigg\rangle_\psi-\bigg\langle\int_{V_{tbp}} \mathrm{d}^3v\:\pdv{}{\psi}\left[\frac{I}{\Omega}\frac{1}{qR}\frac{r}{R}\mathcal{P}(\theta)g^{t,bp}_{0}\right]\bigg\rangle_\psi\\
     +\bigg\langle \left(1+\frac{2I}{\Omega}\pdv{u}{\psi}\right)\int_{V_{tbp}} \mathrm{d}^3v\:\pdv{}{w} \left(\frac{wu}{qR}\frac{r}{R}\sin\theta g^{t,bp}_{0}\right)\bigg\rangle_\psi     
\end{multline}
We show in Appendix \ref{sec: g1} that the $\theta$-dependent piece of $g_1^{t,bp}$ decays for $w\rightarrow\pm\infty$ and, consequently, the first term in \eqref{Gamma g1} vanishes. The $\theta$-dependence of $g^{t,bp}_{0}$ is such that $wg^{t,bp}_0\rightarrow K$ for $w\rightarrow\pm\infty$, with $K$ being the same constant for $w\rightarrow+\infty$ and $w\rightarrow -\infty$. Thus, $\langle  wg^{t,bp}_{0}\sin\theta\rangle_\psi$ does not jump across the trapped--barely passing region, and the second and fourth terms vanish when integrated over $w$. The third term is the only term that does not vanish. Thus, we find
\begin{equation}
    \bigg\langle\int \mathrm{d}^3v\:\Sigma\bigg\rangle_\psi =\pdv{\Gamma}{\psi},
\end{equation}
where
\begin{equation}\label{Gamma g0}
    \Gamma=-\bigg\langle\int_{V_{tbp}} \mathrm{d}^3v\:\frac{I}{\Omega}\frac{1}{qR}\frac{r}{R}\mathcal{P}(\theta)g^{t,bp}_{0}\bigg\rangle_\psi 
\end{equation}
is the ion neoclassical particle flux. We need to calculate $g^{t,bp}_{0}$ to evaluate this integral. \par

Using the same manipulations that we used for the particle conservation equation, we find that the parallel momentum moment of $\mathcal{L}[g^{t,bp}]$ in the trapped--barely passing region is 
\begin{multline}\label{lowest momentum}
     \bigg\langle\int_{V_{tbp}} \mathrm{d}^3v\: m(w-u)\mathcal{L}[g^{t,bp}]\bigg\rangle_\psi \simeq mu\bigg\langle\int_{V_{tbp}} \mathrm{d}^3v\:\pdv{}{\psi}\left[\frac{I}{\Omega}\frac{1}{qR}\frac{r}{R}\mathcal{P}(\theta)g^{t,bp}_{0}\right]\bigg\rangle_\psi\\
     -S\bigg\langle \int_{V_{tbp}} \mathrm{d}^3v\:mw\frac{1}{qR}\frac{r}{R}\pdv{}{w} \left[\mathcal{P}(\theta)g^{t,bp}_{0}\right]\bigg\rangle_\psi.
\end{multline}
Note that the second term is the result of keeping the small difference between $v_\parallel$ and $-u$ in \eqref{passing momentum}. In \eqref{lowest Gamma}, the particle moment of this term did not give a contribution, but here we need to keep it. We can integrate the second term in \eqref{lowest momentum} by parts to relate it to the particle flux \eqref{Gamma g0},
\begin{equation} \label{lowest momentum 2}
     \bigg\langle\int_{V_{tbp}} \mathrm{d}^3v\: m(w-u)\mathcal{L}[g^{t,bp}]\bigg\rangle_\psi \simeq -mu\pdv{\Gamma}{\psi} +mS\frac{\Omega}{I}\Gamma.
\end{equation}
There was no jump term when we integrated by parts because $\langle w g_0^{t,bp}\mathcal{P}(\theta)\rangle_\psi$ does not jump across the trapped--barely passing region for the same reason as argued below \eqref{Gamma g1}.
We can use \eqref{lowest momentum 2} in the parallel momentum relation \eqref{passing momentum} and find the parallel momentum conservation equation
\begin{equation}\label{gammac1}
    \bigg\langle\int \mathrm{d}^3v\:mv_\parallel\Sigma\bigg\rangle_\psi=\pdv{}{\psi}\left(-mu\Gamma\right)-m\frac{\Omega}{I}\Gamma.
\end{equation}
\par
Finally, using again the same manipulations that we used for the particle and parallel momentum conservation equation, we find that energy transport \eqref{Energy total} is described by the equation
\begin{multline}\label{energy 0}
    \bigg\langle\int \mathrm{d}^3v\:\frac{mv^2}{2}\Sigma\bigg\rangle_\psi =-\bigg\langle\int_{V_{tbp}} \mathrm{d}^3v\:\left(\frac{mu^2}{2}+m\mu B\right)\pdv{}{\psi}\left[\frac{I}{\Omega}\frac{1}{qR}\frac{r}{R}\mathcal{P}(\theta)g^{t,bp}_{0}\right]\bigg\rangle_\psi\\
  -S\bigg\langle\int_{V_{tbp}} \mathrm{d}^3v\:muw\frac{1}{qR}\frac{r}{R}\pdv{}{w}\left[\mathcal{P}(\theta)g^{t,bp}_{0}\right]\bigg\rangle_\psi.
\end{multline}
The second term on the right hand side is due to the small difference of $v_\parallel$ and $-u$ in the trapped--barely passing region. We can integrate the second term by parts, because $\langle w \mathcal{P}(\theta)g_0^{t,bp}$ does not jump across the trapped--barely passing region, and relate it to the particle flux,
\begin{multline}\label{energy 1}
    \bigg\langle\int \mathrm{d}^3v\:\frac{mv^2}{2}\Sigma\bigg\rangle_\psi =-\pdv{}{\psi}\bigg\langle\int_{V_{tbp}} \mathrm{d}^3v\:\left(\frac{mu^2}{2}+m\mu B\right)\left[\frac{I}{\Omega}\frac{1}{qR}\frac{r}{R}\mathcal{P}(\theta)g^{t,bp}_{0}\right]\bigg\rangle_\psi\\
    -mu\pdv{u}{\psi}\Gamma +muS\frac{\Omega}{I}\Gamma.
\end{multline}
The energy conservation equation \eqref{energy 1} can be written as
\begin{equation}\label{energy pre}
    \bigg\langle\int \mathrm{d}^3v\:\frac{mv^2}{2}\Sigma\bigg\rangle_\psi =\pdv{Q}{\psi}
    +mu\frac{\Omega}{I}\Gamma,
\end{equation}
with the energy flux
\begin{equation} \label{Q g0}
    Q= -\bigg\langle\int_{V_{tbp}} \mathrm{d}^3v\:\left(\frac{mu^2}{2}+m\mu B\right)\frac{I}{\Omega}\frac{1}{qR}\frac{r}{R}\mathcal{P}(\theta)g^{t,bp}_{0}\bigg\rangle_\psi .
\end{equation}

\subsection{Electron neoclassical transport}
The electron neoclassical transport relations can be derived following the exact same steps as for the ions. The main differences are that the square root of the mass ratio $\sqrt{m_e/m} \ll\epsilon\ll  1$, where $m_e$ is the electron mass, introduces a second small parameter and that electron-ion collisions must be kept in addition to electron-electron collisions. We showed in \cite{trinczek2025a} that this leads to a collision operator where we can replace the term $\mathcal{\sf{M}}_\parallel$ by
\begin{equation}
    \mathcal{\sf{M}}_{\parallel e}=\frac{3}{2}\sqrt{\frac{\pi}{2}}\frac{T_e}{m_e}\frac{\nu_{ee}}{x_e}\left[\Theta(x_e)-\Psi(x_e)+Z\right].
\end{equation}
Here, $T_e$ is the electron temperature, $\nu_{ee}=4\sqrt{\pi}e^4n_e\log\Lambda/(3T_e^{3/2}m_e^{1/2})$, $n_e$ is the electron density and $x_e=v/v_{te}$, where $v_{te}=\sqrt{2T_e/m_e}$ is the electron thermal speed. Note that the definition of $\nu_{ee}$ is consistent with the definition in \cite{trinczek2025a} but differs by a factor of $\sqrt{2}$ from the standard Braginskii definition.
Dropping terms that are small in mass ratio and using that the electron charge is $-e$, we find the electron neoclassical particle flux
\begin{equation}\label{Gammae g0}
    \Gamma_e=-\bigg\langle\int_{V_{tbp}}\mathrm{d}^3v\:\frac{I}{\Omega_e}\frac{1}{qR}\frac{r}{R}\mathcal{P}_e(\theta)g^{t,bp}_{e0}\bigg\rangle_\psi   ,
\end{equation}
where $\Omega_e=-eB/m_ec$ is the electron Larmor frequency and
\begin{equation}
    \mathcal{P}_e(\theta)=\mu B\sin\theta-\frac{eR}{m_er}\pdv{\phi_\theta}{\theta}.
\end{equation}
The electron neoclassical energy flux is
\begin{equation} \label{Qe g0}
    Q_e= -\bigg\langle\int_{V_{tbp}} \mathrm{d}^3v\:m_e\mu B\frac{I}{\Omega_e}\frac{1}{qR}\frac{r}{R}\mathcal{P}_{e}(\theta)g^{t,bp}_{e0}\bigg\rangle_\psi.
\end{equation}
\par
The bootstrap current can also be calculated from the electron distribution function. 
We follow the derivation for the bootstrap current of \cite{trinczek2025a} but generalize it for $\nu_\ast\sim1$. The bootstrap current is defined as
\begin{equation}\label{bootstrap definition}
    j_\parallel^B\equiv Ze\int\mathrm{d}^3v\: v_\parallel f - e\int\mathrm{d}^3v \: v_\parallel f_e=ZenV_\parallel -en_eV_{\parallel } -e\int\mathrm{d}^3v\: v_\parallel g_e
\end{equation}
Quasineutrality reduces this expression to
\begin{equation}\label{bootstrap just ge}
    j_\parallel^B=-e\int\mathrm{d}^3v\: v_\parallel g_{e}.
\end{equation}
Using the same approach as \cite{trinczek2025a}, we can use the Spitzer-Härm function in \eqref{bootstrap just ge}. The Spitzer-Härm function is defined to satisfy the property
\begin{equation}\label{Spitzer property}
    v_\parallel f_{Me}=C_e[f_{e,SH}].
\end{equation}
The Spitzer-Härm function is of the form
\begin{equation}\label{Spitzer function}
    f_{e,SH}=\frac{v_\parallel}{\sqrt{2}\nu_{ee}}f_{Me}A_{SH}\!\left(x_e^2\right),
\end{equation}
where
\begin{equation}
    A_{SH}=\sum_i a_i L_i^{3/2}(x_e^2).
\end{equation}
Here, $L_i^{3/2}(x_e^2)$ are Laguerre polynomials and $a_i$ are coefficients determined by \eqref{Spitzer property}. The first three coefficients for $Z=1$ are $a_0=-1.975$, $a_1=0.558$, $a_2=0.015$. We use \eqref{Spitzer property} in \eqref{bootstrap just ge} to write the flux surface averaged bootstrap current as
\begin{equation}\label{self adjoint}
    \langle j_\parallel^B\rangle_\psi=-e\Bigg\langle \int\mathrm{d}^3v\: \frac{g_{e}}{f_{Me}}C_e[f_{e,SH}]\Bigg\rangle_\psi =-e\Bigg\langle \int\mathrm{d}^3v\: \frac{f_{e,SH}}{f_{Me}}C_e[g_{e}]\Bigg\rangle_\psi,
\end{equation}
where we have used the self adjointness of the collision operator. The bootstrap current can be expressed as a moment of the collision operator,
\begin{equation}
    \langle j_\parallel^B\rangle_\psi\simeq-\Bigg\langle \int\mathrm{d}^3v\: \frac{e}{\sqrt{2}\nu_{ee}} v_\parallel A_{SH} C_e[g_e]\Bigg\rangle_\psi.
\end{equation}
At this point, we can employ \eqref{DK in L},
\begin{multline}
    \langle j_\parallel^B\rangle_\psi\simeq-\Bigg\langle \int_{V_{tbp}}\mathrm{d}^3v\: \frac{e}{\sqrt{2}\nu_{ee}} v_\parallel A_{SH} \mathcal{L}[g_e^{t,bp}]\Bigg\rangle_\psi\\
    \simeq \Bigg\langle \int_{V_{tbp}}\mathrm{d}^3v\: \frac{e}{\sqrt{2}\nu_{ee}} w A_{SH} \frac{1}{qR}\frac{r}{R}\pdv{}{w} \left[\mathcal{P}_e(\theta)g^{t,bp}_{e0}\right]\Bigg\rangle_\psi.
\end{multline}
Upon integration by parts, where we use that $\langle w\mathcal{P}_e(\theta)g_{e0}^{t,bp}\rangle_\psi$ does not jump across the trapped--barely passing region as argued below \eqref{Gamma g1} and $x_e^2\simeq m_e\mu B/T_e$, the bootstrap current is given by
\begin{equation}\label{bootstrap integral}
    \langle j_\parallel^B\rangle_\psi=-\bigg\langle\int_{V_{tbp}}\mathrm{d}^3v\:  \frac{e}{\sqrt{2}\nu_{ee}}A_{SH}\frac{1}{qR}\frac{r}{R}\mathcal{P}_e(\theta)g_{e0}^{t,bp}\bigg\rangle_\psi.
\end{equation}
\par

All results up to this point are valid for $\nu_\ast\sim 1$ and to lowest order in $\sqrt{\epsilon}$. To calculate the particle flux and the energy flux, it is necessary to distinguish between the different collisionality regimes. The calculation for the banana regime, $\nu_\ast\ll1$, was carried out in \cite{Trinczek2023}. We present the derivation for the plateau regime, $\nu_\ast\gg 1$, in the next section.

\section{Plateau regime}\label{sec: Plateau}
The plateau regime is defined by
\begin{equation}
    1\ll \nu_\ast\ll\epsilon^{-3/2} \quad \mbox{and\ }\quad \epsilon \ll1.
\end{equation}
The collision frequency is small compared to the transit time of freely passing particles but big enough that trapped and barely passing particles collide many times before they can complete their orbits. Hence, we will refer to the particles with small $w$ as being in a collisional layer of a width that we determine in a few lines, and refrain from distinguishing trapped and barely passing particles. The distribution function in this collisional layer is called $g^l$ and replaces the distribution function $g^{t,bp}$. The width of the collisional layer in $w$ is determined by imposing that the effective collision frequency $\nu v_{t}^2/w^2$ be comparable to the time it takes a particle to complete one full poloidal turn,
\begin{equation}
    \frac{w}{qR}\sim\nu\frac{v^2_t}{w^2}
\end{equation}
and hence the width of the collisional layer is
 \begin{equation}
 \frac{w}{v_{t}}\sim\left(\frac{qR\nu}{v_{t}}\right)^{1/3}\sim\epsilon^{1/2}\nu_\ast^{1/3}.
\end{equation}

The lowest order correction to the distribution function in the Plateau regime is of the size $g^l_{0}/f_M\sim\epsilon^{1/2}/\nu_\ast^{1/3}\ll 1$. 
\subsection{Ion transport in the plateau regime}
In order to determine $g^l_0$ in the plateau regime, we can use that the second term on the left hand side in \eqref{General Jump Condition}, the term proportional to $\partial g_0^l/\partial w$, is smaller than other terms by $1/\nu_\ast^{2/3}\ll1$. Thus, \eqref{General Jump Condition} simplifies to
\begin{equation}\label{Jump Condition Plateau}
    \pdv{}{\theta}\left(\frac{w}{qR}g^l_{0}\right)-\mathsf{M}_\parallel \pdv[2]{g^l_{0}}{w}=\frac{I}{\Omega}\mathcal{P}(\theta)\frac{r}{R}\frac{1}{qR} \mathcal{D}f_M,
\end{equation}
This equation is analytically solvable if the $\theta$-dependence of $u$ is neglected as small in $\epsilon$ and the form of the potential is taken to be
\begin{equation} \label{potential assumption plateau}
    \phi_\theta=\phi_c(\psi)\cos\theta+\phi_s(\psi)\sin\theta.
\end{equation}
This choice will be justified for concentric circular flux surfaces in the calculation of the electric potential in section \ref{sec: potential plateau}. Introducing the amplitudes
\begin{align}
   A_S\equiv u^2+\mu B-\frac{Ze\phi_c R}{mr}, && A_C\equiv\frac{Ze\phi_sR}{m r},
\end{align}
we can write
\begin{equation}
    \mathcal{P}(\theta)=A_S\sin\theta+A_C\cos\theta
\end{equation}
in \eqref{Jump Condition Plateau}. The full derivation of the solution to \eqref{Jump Condition Plateau} is in Appendix \ref{sec: g0}, which follows \cite{su1968}.
The solution is
\begin{equation}\label{gl}
g^l_{0}=\left[A_SF_s(\xi,\theta)+A_CF_c(\xi,\theta)\right]\frac{I}{\Omega} \frac{\mathcal{D}}{v_\textrm{ref}}\frac{r}{R} f_M(w=0),
\end{equation}
where $\xi=w/v_\textrm{ref}$, $v_\textrm{ref}=(qR\mathsf{M}_\parallel)^{1/3}$,
\begin{equation}\label{Fs}
    F_s(\xi,\theta)=\int_0^\infty\mathrm{d}p \:\exp\left(-\frac{p^3}{3}\right)\sin(\theta-p\xi) 
\end{equation}
and
\begin{equation}\label{Fc}
    F_c(\xi,\theta)=\int_0^\infty\mathrm{d}p\:\exp\left(-\frac{p^3}{3}\right)\cos(\theta-p\xi).
\end{equation}
Importantly, $\langle g^l_{0}\rangle_\tau=0$ and $g^l_{0}$ matches \eqref{theta dependent gp} for $\xi\rightarrow\infty$, a result demonstrated in Appendix \ref{sec: g0}. We also need the integral
\begin{equation}\label{int gl}
 \int\text{d}w\:g^l_{0}=\frac{\pi I}{\Omega}\frac{r}{R}\mathcal{D}\left(A_C\cos\theta+A_S\sin\theta\right)f_M(w=0)
\end{equation}
to calculate the particle flux $\Gamma$ in \eqref{Gamma g0}. The derivation of \eqref{int gl} is presented in Appendix \ref{sec: g0}.\par
We can use \eqref{int gl} in \eqref{Gamma g0} to find the ion neoclassical particle flux
\begin{equation}
    \Gamma=-\frac{BI^2\pi^2}{qR\Omega^2}\left(\frac{r}{R}\right)^2\int\text{d}\mu\left(A_S^2+A_C^2\right)\mathcal{D}f_M(w=0),
\end{equation}
where we also carried out the flux surface average.
The integration over $\mu$ gives
\begin{multline}\label{Gamma}
    \Gamma=-\frac{\sqrt{\pi}}{4}\frac{n}{qR}\frac{I^2}{\Omega^2}\left(\frac{r}{R}\right)^2\left(\frac{2T}{m}\right)^{3/2}\exp\left(-\frac{m(u+V_\parallel)^2}{2T}\right)\\
    \times\Bigg\lbrace\frac{1}{2}\left[\left(\frac{mu^2}{T}-\frac{ZeR\phi_c}{Tr}\right)^2+\left(\frac{ZeR\phi_s}{Tr}\right)^2\right]\mathcal{D}_{-3/2} +\left(\frac{mu^2}{T}-\frac{ZeR}{Tr}\phi_c\right)\mathcal{D}_{-1/2}+\mathcal{D}_{1/2}\Bigg\rbrace.
\end{multline}
Here, we have introduced the notation,
\begin{equation}
    \mathcal{D}_l\equiv \pdv{}{\psi}\ln p-\frac{m(u+V_\parallel)}{T}\left(\pdv{V_\parallel}{\psi}-\frac{\Omega}{I}\right)+\left[\frac{m(u+V_\parallel)^2}{2T}+l\right]\pdv{}{\psi}\ln T,
\end{equation}
where $l$ is a rational number.\par
The energy flux $Q$ in \eqref{Q g0} is
\begin{equation}\label{K}
    Q=-\frac{BTI^2\pi^2}{qR\Omega^2}\left(\frac{r}{R}\right)^2\int\text{d}\mu\:\left(\frac{mu^2}{2T}+\frac{m \mu B}{T}\right)\left(A_S^2+A_C^2\right)\mathcal{D}f_M(w=0),
\end{equation}
which can be calculated to give
\begin{multline}\label{Q}
       Q=-\frac{3\sqrt{\pi}}{4}\frac{nT}{qR}\frac{I^2}{\Omega^2 }\left(\frac{r}{R}\right)^2
       \left(\frac{2T}{m}\right)^{3/2}\exp\left(-\frac{m(u+V_\parallel)^2}{2T}\right)\\
       \times\Bigg\lbrace\frac{mu^2}{12T}\Bigg[\left(\frac{mu^2}{T}-\frac{Ze\phi_cR}{Tr}\right)^2+\left(\frac{Ze\phi_sR}{Tr}\right)^2\Bigg]\mathcal{D}_{-3/2}\\
        +\frac{1}{6}\left[\left(\frac{2mu^2}{T}-\frac{Ze\phi_c R}{Tr}\right)\left(\frac{mu^2}{T}-\frac{Ze\phi_c R}{Tr}\right) + \left(\frac{Ze\phi_s R}{Tr}\right)^2\right]\mathcal{D}_{-1/2}\\
        +\left(\frac{5}{6}\frac{mu^2}{T}-\frac{2}{3}\frac{Ze\phi_cR}{Tr}\right)\mathcal{D}_{1/2}+\mathcal{D}_{3/2}\Bigg\rbrace.
    \end{multline}
\par In the limit of weak gradients, the poloidal variation vanishes, and the neoclassical ion particle flux \eqref{Gamma} and energy flux \eqref{Q} agree with the results of weak gradient neoclassical theory, as shown in Appendix \ref{sec: weak grad}.\par 
We compare the results in \eqref{Gamma} and \eqref{Q} to previous work on strong gradient neoclassical transport in the plateau regime by \cite{Seol2012} and \cite{Pusztai2010} in Appendix \ref{sec: Comparison Peter/Shaing}, and find that our results disagree due to inconsistencies in previous work.

\subsection{Electron transport in the plateau regime}
The electron distribution function $g^l_{e0}$ can be derived following the exact same steps as for the ion distribution function. The result is
\begin{equation}\label{ge0}
g^l_{e0}=\left[A_{Se}F_s(\xi_e,\theta)+A_{Ce}F_c(\xi_e,\theta)\right]\frac{I}{\Omega_e} \frac{\mathcal{D}_e}{v_\textrm{ref,e}}\frac{r}{R} f_{Me}(v_\parallel=0),
\end{equation}
where
$\xi_e=v_\parallel/v_\textrm{ref,e}$, $v_\textrm{ref,e}=(qR\mathsf{M}_{\parallel e})^{1/3}$,
\begin{align}
    A_{Se}=\mu B+\frac{e\phi_c R}{m_e r},&&A_{Ce}=-\frac{e\phi_sR}{m_er},
\end{align}
\begin{equation}
    \mathcal{D}_e=\pdv{}{\psi}\ln{p_e} +\frac{m_e(u+ V_\parallel)}{T}\frac{\Omega_e}{I} +\left(\frac{m_e\mu B}{T_e}-\frac{5}{2}\right)\pdv{}{\psi}\ln{T_e}
\end{equation}
and $p_e=n_eT_e$ is the electron pressure.
We can use the distribution function \eqref{ge0} in \eqref{Gammae g0} and integrate to find
\begin{multline}\label{Gammae}
    \Gamma_e=-\frac{\sqrt{\pi}}{4}\frac{n_e}{qR}\frac{I^2}{\Omega_e^2}\left(\frac{r}{R}\right)^2\left(\frac{2T_e}{m_e}\right)^{3/2}\bigg\lbrace\frac{1}{2}\left[\left(\frac{e\phi_cR}{T_er}\right)^2+\left(\frac{e\phi_s R}{T_e r}\right)^2\right]\mathcal{D}_{e,-3/2}\\
    +\frac{e\phi_c R}{T_e r}\mathcal{D}_{e,-1/2}+\mathcal{D}_{e,1/2}\bigg\rbrace,
\end{multline}
where
\begin{equation}
    \mathcal{D}_{e,l}\equiv \pdv{}{\psi}\ln p_e+\frac{m_e(u+V_\parallel)}{T_e}\frac{\Omega_e}{I}+l\pdv{}{\psi}\ln T_e.
\end{equation}
Similarly, the electron energy flux \eqref{Qe g0} yields
\begin{multline}\label{Qe}
    Q_e=-\frac{3\sqrt{\pi}}{4}\frac{n_eT_e}{qR}\frac{I^2}{\Omega_e^2}\left(\frac{r}{R}\right)^2\left(\frac{2T_e}{m_e}\right)^{3/2}\bigg\lbrace\frac{1}{6}\left[\left(\frac{e\phi_cR}{T_er}\right)^2+\left(\frac{e\phi_s R}{T_e r}\right)^2\right]\mathcal{D}_{e,-1/2}\\
    +\frac{2}{3}\frac{e\phi_c R}{T_e r}\mathcal{D}_{e,1/2}+\mathcal{D}_{e,3/2}\bigg\rbrace.
\end{multline}
In the limit of weak gradients, the neoclassical electron particle flux \eqref{Gammae} and energy flux \eqref{Qe} agree with the results of weak gradient neoclassical theory, as shown in Appendix \ref{sec: weak grad}.
\par
We can insert the electron distribution function \eqref{ge0} in \eqref{bootstrap integral} to calculate the bootstrap current. We take the integral \eqref{bootstrap integral} by using properties of the Laguerre polynomials. The derivation is shown in detail in Appendix \ref{sec: bootstrap Appendix}. The final result for the bootstrap current reads
\begin{multline}\label{bootstrap}
    \langle j_\parallel^B\rangle_\psi= -\frac{1}{8}\sqrt{\frac{\pi}{2}}\frac{en_e}{\nu_{ee}qR}\frac{I}{\Omega_e}\left(\frac{r}{R}\right)^2\left(\frac{2T_e}{m_e}\right)^{3/2}\bigg\lbrace \left[\left(\frac{e\phi_cR}{T_er}\right)^2+\left(\frac{e\phi_sR}{T_er}\right)^2\right]\\
    \times\sum_{i=0} \frac{2a_i\Gammaf(i+\frac{3}{2})}{\sqrt{\pi}\Gammaf(1+i)}\mathcal{D}_{e,\alpha_i}    +2\frac{e\phi_cR}{T_e r}\left[a_0\mathcal{D}_{e,-\frac{1}{2}}+\sum_{i=1}\frac{a_i\Gammaf(i+\frac{1}{2})}{\sqrt{\pi}\Gammaf(1+i)}\mathcal{D}_{e,\beta_i}\right]\\
    +\left[2a_0\mathcal{D}_{e,1/2}-a_1\mathcal{D}_{e,13/2}-\sum_{i=2} \frac{a_i\Gammaf(i-\frac{1}{2})}{\sqrt{\pi}\Gammaf(1+i)}  \mathcal{D}_{e,\gamma_i} \right]\bigg\rbrace,
\end{multline}
where
\begin{align}
    \alpha_i=-\frac{3/2+5i}{1+2i},&&\beta_i=\frac{1-10i}{4i-2},&&\gamma_i=-\frac{(3+10i)(i-1/2)}{4}.
\end{align}
Note that $\Gammaf$ here is the Euler Gamma function and not the particle flux. The result agrees with the weak gradient limit in Appendix \ref{sec: weak grad} and we compare it to previous results by \cite{Pusztai2010} in Appendix \ref{sec: Comparison Peter/Shaing}.

\subsection{Electric potential} \label{sec: potential plateau}
The calculation of the electric potential in the plateau regime follows the derivation in the banana regime in \cite{Trinczek2023}. We assume a Boltzmann response for the electrons such that quasineutrality reads
\begin{equation}\label{Boltzmann}
    Zn_\theta=\frac{en_e}{T_e}\phi_\theta.
\end{equation}
\par The poloidal density variation as derived in (4.56) in \cite{Trinczek2023} is 

\begin{multline}\label{ntheta2}
    n_\theta=\int\mathrm{d}\mu\int\mathrm{d}w\: 2\pi B (g^l-\langle g^l\rangle_\psi)+\int\mathrm{d}\mu\:\left[\text{PV}\!\int\mathrm{d}v_\parallel\: 2\pi B(g^p-\langle g^p\rangle_\psi)\right]\\
    \equiv n_\theta^l+n_\theta^p,
\end{multline}
where the integration is performed over both the collisional layer and the freely passing region of velocity space, and hence one must calculate the contributions from $g^l$ and $g^p$ separately.
The freely passing particle contribution is the same as in the banana regime, and it requires a principal value integral around $v_\parallel=-u$ because $g^p$ diverges there as $1/(v_\parallel +u)$ - see \eqref{theta dependent gp}. The passing particle contribution was calculated by \cite{Trinczek2023} and yields
\begin{multline}\label{poloidal density 2}
    n_\theta^p=- n\frac{Ir}{\Omega R}\Bigg\lbrace \sqrt{\frac{2T}{m}}J \bigg[\left(\frac{mV_\parallel^2}{T}\cos\theta+\cos\theta -\frac{Ze\phi_\theta R}{Tr}\right)\left(\pdv{}{\psi}\ln p-\frac{3}{2}\pdv{}{\psi}\ln T\right)\\
   +\cos\theta\pdv{}{\psi}\ln T \bigg]
    +\left[1-2\sqrt{\frac{m}{2T}}(V_\parallel+u)J\right]\Bigg\lbrace(V_\parallel-u)\cos\theta\left(\pdv{}{\psi}\ln p-\frac{3}{2}\pdv{}{\psi}\ln T\right)\\
    -(V_\parallel+u)\Bigg[\left(\frac{mV_\parallel^2}{2T}+\frac{1}{2}\right)\cos\theta -\frac{Ze\phi_\theta R}{2Tr}\Bigg]\pdv{}{\psi}\ln T+\left(\pdv{V_\parallel}{\psi}-\frac{\Omega}{I}\right)\\
    \times\left[\left(\frac{m u^2}{T}+1-\frac{m(V_\parallel+u)^2}{T}\right)\cos\theta-\frac{Ze\phi_\theta R}{Tr}\right]\Bigg\rbrace\\
        +\left[1+2\frac{m(V_\parallel+u)^2}{2T}-4\left(\frac{m}{2T}\right)^{3/2}(V_\parallel+u)^3J\right]\cos\theta\left(\pdv{V_\parallel}{\psi}-\frac{\Omega}{I}+\frac{V_\parallel-u}{2}\pdv{}{\psi}\ln T\right)\Bigg\rbrace\\
        -2n\frac{r}{R}\cos\theta,
\end{multline}
where $J$ is defined as
\begin{equation}\label{J}
    J=\frac{\sqrt{\pi} }{2}\exp\left(-\frac{m(u+V_\parallel)^2}{2T}\right)\erfi\left(\sqrt{\frac{m}{2T}}(u+V_\parallel)\right).
\end{equation}
Note that the dependence on $\cos \theta$ in this expression is coming from the magnetic field strength.\par
In the banana regime, the contribution from trapped and barely passing particles to the density vanishes because the $\theta$-dependent part of the distribution function is odd in $w$. In the plateau regime, the contribution from the particles in the collisional layer does not vanish to lowest order. The term $\int\text{d}\mu \int\text{d}w\text{ }2\pi B g^l_{0}$ needs to be evaluated, whereas $\langle g^l_{0}\rangle_\tau$ vanishes. The velocity integration of $g^l_{0}$ can be performed using the solution for $g^l_{0}$, \eqref{gl}, the relation \eqref{int gl}, and assuming \eqref{potential assumption plateau},
\begin{multline}\label{poloidal density trapped}
       n^l_\theta\simeq \int\text{d}\mu\int\text{d}w \: 2\pi B g^l_{0}= n\sqrt{\frac{m}{T}}\sqrt{\frac{\pi}{2}}\exp\left(-\frac{m(u+V_\parallel)^2}{2T}\right)\\
       \times\Bigg\lbrace\frac{I}{\Omega}\Bigg[\frac{Ze}{m}\phi_s\cos\theta
       +\left(u^2\frac{r}{R}-\frac{Ze}{m}\phi_c\right)\sin \theta\Bigg] \mathcal{D}_{-3/2}+\frac{ITr}{\Omega m R}\sin\theta\mathcal{D}_{-1/2}\Bigg\rbrace.
\end{multline}

\par Combining \eqref{poloidal density 2} and \eqref{poloidal density trapped} in \eqref{Boltzmann} yields an equation for the poloidally varying component of the electric potential $\phi_\theta=\phi_c(\psi)\cos\theta+\phi_s(\psi)\sin\theta$. The terms proportional to $\cos\theta$ give
\begin{multline}\label{phic plateau}
    \Bigg\lbrace \frac{en_e}{T_e}-\frac{Z^2enI}{T\Omega}\bigg[\sqrt{\frac{2T}{m}}J\left(\pdv{}{\psi}\ln p-\frac{3}{2}\pdv{}{\psi} \ln T\right)+\left(1-2\sqrt{\frac{m}{2T}}(V_\parallel+u) J\right)\Big(\pdv{V_\parallel}{\psi}-\frac{\Omega}{I}\\
    -\frac{(V_\parallel+u)}{2}\pdv{}{\psi}\ln T\Big)\bigg] \Bigg\rbrace\phi_c=-Zn\frac{Ir}{\Omega R}\Bigg\lbrace \sqrt{\frac{2T}{m}}J\bigg[\left(\frac{mV_\parallel^2}{T}+1\right)\\
    \times \left(\pdv{}{\psi}\ln p-\frac{3}{2}\pdv{}{\psi} \ln T\right)+\pdv{}{\psi}\ln T\bigg] +\left[1-2\sqrt{\frac{m}{2T}}(V_\parallel+u) J\right]\\
    \times\bigg[(V_\parallel-u)\left(\pdv{}{\psi}\ln p-\frac{3}{2}\pdv{}{\psi}\ln T\right)+\left(\pdv{V_\parallel}{\psi}-\frac{\Omega}{I}\right)\left(\frac{mu^2}{T}+1-\frac{m(V_\parallel+u)^2}{T}\right)\\
    -\frac{V_\parallel+u}{2}\left(\frac{mV_\parallel^2}{T}+1\right)\pdv{}{\psi}\ln T\bigg]\\
    +\left[1+\frac{m}{T}(V_\parallel+u)^2-4\left(\frac{m}{2T}\right)^{3/2}\left(V_\parallel+u\right)^3 J\right]\left(\pdv{V_\parallel}{\psi}-\frac{\Omega}{I}+\frac{V_\parallel-u}{2}\pdv{}{\psi}\ln T\right)\Bigg\rbrace\\
    +Zn\sqrt{\frac{m}{T}}\sqrt{\frac{\pi}{2}}\exp\left(-\frac{m(u+V_\parallel)^2}{2T}\right)\frac{Ze}{m}\phi_s  \mathcal{D}_{-3/2}-2Zn\frac{r}{R},
\end{multline}
and all terms proportional to $\sin\theta$ give
\begin{multline}\label{phis plateau}
    \Bigg\lbrace \frac{en_e}{T_e}-\frac{Z^2enI}{T\Omega}\bigg[\sqrt{\frac{2T}{m}}J\left(\pdv{}{\psi}\ln p-\frac{3}{2}\pdv{}{\psi} \ln T\right)+\left(1-2\sqrt{\frac{m}{2T}}(V_\parallel+u)J\right)\\
     \times\Big(\pdv{V_\parallel}{\psi}-\frac{\Omega}{I}-\frac{V_\parallel+u}{2}\pdv{}{\psi}\ln T\Big)\bigg] \Bigg\rbrace\phi_s=Zn\sqrt{\frac{m}{T}}\sqrt{\frac{\pi}{2}}\exp\left(-\frac{m(u+V_\parallel)^2}{2T}\right) \\
     \times\Bigg\lbrace\frac{I}{\Omega}\left(u^2\frac{r}{R}-\frac{Ze}{m}\phi_c\right)\mathcal{D}_{-3/2} +\frac{TIr}{m\Omega R}\mathcal{D}_{-1/2}\Bigg\rbrace.
\end{multline}
Equations \eqref{phic plateau} and \eqref{phis plateau} form a linear system that can be solved for $\phi_c$ and $\phi_s$.
This result is consistent with our choice of $\phi_\theta$ in \eqref{potential assumption plateau}. 
\par The usual neoclassical results for the electric potential can be retrieved by taking the limit $u\ll v_{t}$,  $V_\parallel \ll v_{t}$ and weak gradients. In this weak gradient limit, equations \eqref{phic plateau} and \eqref{phis plateau} reduce to
\begin{equation}
   \phi_c=0
\end{equation}
and
\begin{equation} \label{Phis wg}
    \left( \frac{en_e}{T_e}+\frac{Z^2en}{T}\right)\phi_s =Zn\sqrt{\frac{T}{m}}\sqrt{\frac{\pi}{2}}
 \frac{Ir}{\Omega R}\left[\pdv{}{\psi}\ln p+\frac{m(u+V_\parallel)}{T}\frac{\Omega}{I}-\frac{1}{2}\pdv{}{\psi}\ln T\right].
\end{equation}
The vanishing parallel friction force in the weak gradient limit gives a relation for the mean parallel flow \eqref{Vwg}, which can be used to simplify \eqref{Phis wg} to
\begin{equation}
    \left( \frac{en_e}{T_e}+\frac{Z^2en}{T}\right)\phi_s =-Zn\sqrt{\frac{T}{m}}\sqrt{\frac{\pi}{2}}
 \frac{Ir}{\Omega R}\pdv{}{\psi}\ln T.
\end{equation}
A non-vanishing up-down asymmetry is consistent with the results by \cite{hinton1973}.

\section{Case study}\label{sec: Case study}
For a given set of input profiles for $n$, $T$, $T_e$ and $V_\parallel$, we can calculate the transport fluxes $\Gamma$, $\Gamma_e$, $Q$ and $Q_e$, the bootstrap current $\langle j_\parallel^B\rangle_\psi$, as well as the poloidal variation amplitudes $\phi_c$ and $\phi_s$. 
We introduce normalised quantities, which are denoted by a bar,
\begin{align}
    \bar{\Gamma}=\frac{\Gamma}{n_0\frac{2T_0}{m}\frac{I\epsilon^2}{qR\Omega}}, && \bar{\gamma}=\frac{\bigg\langle\int \mathrm{d}^3v\:mv_\parallel\Sigma\bigg\rangle_\psi}{n_0 2T_0\frac{\epsilon^2}{qR}}, && \bar{Q}=\frac{Q}{n_0T_0\frac{2T_0}{m}\frac{I\epsilon^2}{qR\Omega}},
\end{align}
\begin{align}
    \bar{\Gamma}_e=\frac{\Gamma_e}{Zn_0\frac{2T_0}{m_e}\frac{I\epsilon^2}{qR\abs{\Omega_e}}\sqrt{\frac{m_e}{m}}},  && \bar{Q}_e=\frac{Q_e}{Zn_0T_0\frac{2T_0}{m_e}\frac{I\epsilon^2}{qR\abs{\Omega_e}}\sqrt{\frac{m_e}{m}}}, && \bar{j}^B=\frac{\langle j_\parallel^B\rangle_\psi}{n_0\frac{2T_0}{m_e}\epsilon^2 \frac{e}{qR\nu_{ee,0}}\sqrt{\frac{m_e}{m}}},
\end{align}
where $T_0$ and $n_0$ are the values of density and temperature at a given point. Furthermore, we define
\begin{align}
    \bar{u}\equiv\sqrt{\frac{m}{2T_0}}u,&&\bar{V}\equiv\sqrt{\frac{m}{2T_0}}V_\parallel, && \bar{\phi}_c\equiv\frac{ZeR\phi_c}{T_0r},&&\bar{\phi}_s\equiv\frac{ZeR\phi_s}{T_0r},&&\bar{E}_r\equiv-\bar{u},
\end{align}
\begin{align}
    \bar{T}\equiv\frac{T}{T_0},&&\bar{T}_e\equiv\frac{T_e}{T_0},&&\bar{n}\equiv\frac{n}{n_0},&&\bar{n}_e\equiv\frac{n_e}{Zn_0}&&\pdv{}{\bar{\psi}}\equiv\frac{I}{\Omega}\sqrt{\frac{2T_0}{m}}\pdv{}{\psi},
\end{align}
\begin{equation}
    \bar{\mathcal{D}}_{e,l}\equiv \frac{I}{\Omega}\sqrt{\frac{2T_0}{m}} \mathcal{D}_{e,l}= \pdv{}{\bar{\psi}}\ln \bar{p}_e-2\frac{\bar{u}+\bar{V}}{Z\bar{T}_e}+l\pdv{}{\bar{\psi}}\ln\bar{T}_e,
\end{equation}
and
\begin{equation}
    \bar{\mathcal{D}}_l\equiv \frac{I}{\Omega}\sqrt{\frac{2T_0}{m}} \mathcal{D}_l= \pdv{}{\bar{\psi}}\ln \bar{p}-2\frac{\bar{u}+\bar{V}}{\bar{T}}\left(\pdv{\bar{V}}{\bar{\psi}}-1\right)+\left[\frac{(\bar{u}+\bar{V})^2}{\bar{T}}+l\right]\pdv{}{\bar{\psi}}\ln\bar{T},
\end{equation}
where $l$ is a rational number.\par
Using these definitions, the neoclassical ion particle flux \eqref{Gamma} is
\begin{multline}\label{HolyGamma}
    \bar{\Gamma}=-\frac{\sqrt{\pi}\bar{n}\bar{T}^{3/2}}{4}\exp\left[-\frac{(\bar{u}+\bar{V})^2}{\bar{T}}\right]\Bigg\lbrace2\left[\left(\frac{\bar{u}^2}{\bar{T}}-\frac{\bar{\phi}_c}{2\bar{T}}\right)^2+\left(\frac{\bar{\phi}_s}{2\bar{T}}\right)^2\right]\bar{\mathcal{D}}_{-3/2}\\
    +2\left(\frac{\bar{u}^2}{\bar{T}}-\frac{\bar{\phi}_c}{2\bar{T}}\right) \bar{\mathcal{D}}_{-1/2}+\bar{\mathcal{D}}_{1/2}\Bigg\rbrace.
\end{multline}
The parallel momentum equation \eqref{gammac1} can be written as
\begin{equation}\label{Holygamma}
    \bar{\gamma}=-\bar{u}\pdv{}{\bar{\psi}}\bar{\Gamma}-S\bar{\Gamma},
\end{equation}
and the neoclassical ion energy flux \eqref{Q} becomes
\begin{multline}\label{HolyQ}
     \bar{Q}=-\frac{3\sqrt{\pi}\bar{n}\bar{T}^{5/2}}{4}\exp\left[-\frac{(\bar{u}+\bar{V})^2}{\bar{T}}\right]\Bigg\lbrace \frac{2\bar{u}^2}{3\bar{T}} \left[\left(\frac{\bar{u}^2}{\bar{T}}-\frac{\bar{\phi}_c}{2\bar{T}}\right)^2+\left(\frac{\bar{\phi}_s}{2\bar{T}}\right)^2\right]\bar{\mathcal{D}}_{-3/2}\\
    +\frac{2}{3}\left[\left(2\frac{\bar{u}^2}{\bar{T}}-\frac{\bar{\phi}_c}{2\bar{T}}\right)\left(\frac{\bar{u}^2}{\bar{T}}-\frac{\bar{\phi}_c}{2\bar{T}}\right)+\left(\frac{\bar{\phi}_s}{2\bar{T}}\right)^2\right] \bar{\mathcal{D}}_{-1/2}+\left(\frac{5}{3} \frac{\bar{u}^2}{\bar{T}}-\frac{2}{3}\frac{\bar{\phi}_c}{\bar{T}} \right)\bar{\mathcal{D}}_{1/2}+\mathcal{D}_{3/2}\Bigg\rbrace.
\end{multline}
The neoclassical electron particle flux \eqref{Gammae} in normalised variables reads
\begin{equation}
    \bar{\Gamma}_e=-\frac{Z\sqrt{\pi}\bar{n}_e\bar{T}_e^{3/2}}{4}\Bigg\lbrace\frac{1}{2}\left[\left(\frac{\bar{\phi}_c}{Z\bar{T}_e}\right)^2+\left(\frac{\bar{\phi}_s}{Z\bar{T}_e}\right)^2\right]\bar{\mathcal{D}}_{e,-3/2}    +\frac{\bar{\phi}_c}{Z\bar{T}_e}\bar{\mathcal{D}}_{e,-1/2}+\bar{\mathcal{D}}_{e,1/2}\Bigg\rbrace.
\end{equation}
The neoclassical electron energy flux \eqref{Qe} is
\begin{multline}\label{HolyQe}
     \bar{Q}_e=-\frac{3\sqrt{\pi}Z\bar{p}_e\bar{T}_e^{3/2}}{4}\Bigg\lbrace \frac{1}{6}
    \left[\left(\frac{\bar{\phi}_c}{Z\bar{T}_e}\right)^2+\left(\frac{\bar{\phi}_s}{Z\bar{T}_e}\right)^2\right] \bar{\mathcal{D}}_{e,-1/2}+\frac{2}{3}
    \frac{\bar{\phi}_c }{Z\bar{T}_e} \bar{\mathcal{D}}_{e,1/2}+\mathcal{D}_{e,3/2}\Bigg\rbrace.
\end{multline}
The bootstrap current \eqref{bootstrap} in normalised variables reads
\begin{multline}\label{bootstrap_norm}
    \bar{j}^B= \frac{Z}{8}\sqrt{\frac{\pi}{2}}\bar{T}_e^3\bigg\lbrace \left[\left(\frac{\bar{\phi}_c}{Z\bar{T}_e}\right)^2+\left(\frac{\bar{\phi}_s}{Z\bar{T}_e}\right)^2\right]\sum_{i=0} \frac{2a_i\Gammaf(\frac{3}{2}+i)}{\sqrt{\pi}\Gammaf(1+i)}\bar{\mathcal{D}}_{e,\alpha_i}\\
    +2\frac{\bar{\phi}_c}{Z\bar{T}_e}\left[a_0\bar{\mathcal{D}}_{e,-\frac{1}{2}} +\sum_{i=1}\frac{a_i\Gammaf(\frac{1}{2}+i)}{\sqrt{\pi}\Gammaf(1+i)}\bar{\mathcal{D}}_{e,\beta_i}\right]\\
    +\left[2a_0\bar{\mathcal{D}}_{e,1/2}-a_1\bar{\mathcal{D}}_{e,13/2}-\sum_{i=2} \frac{a_i\Gammaf(i-\frac{1}{2})}{\sqrt{\pi}\Gammaf(1+i)}  \bar{\mathcal{D}}_{e,\gamma_i} \right]\bigg\rbrace.
\end{multline}
The amplitudes $\bar\phi_c$ and $\bar\phi_s$ that describe the poloidal variation of the potential are given in Appendix \ref{sec: Transport equations}. With the set of transport equations \eqref{HolyGamma}-\eqref{bootstrap_norm}, one can determine the fluxes and poloidal variation amplitudes for a given set of profiles for the radial electric field, the density, the temperatures and the mean parallel flow.\par

We will use an assumption in addition to the input profiles and the transport relations \eqref{HolyGamma}-\eqref{bootstrap_norm} to relate the radial electric field to other quantities. We compare two different approaches. \par 

The first approach is to assume radial force balance in the pedestal,
\begin{equation}\label{radialForceBalance}
    Zen\pdv{\Phi}{\psi}+\pdv{p}{\psi}=0.
\end{equation}
This pressure balance equation states that the radial electric field is set mostly by the pressure gradient. Experimental observations of the pedestal support this assumption \citep{mcdermott2009, Viezzer2013}. However, \eqref{radialForceBalance} does not hold in the core, so we do not expect to recover weak gradient neoclassical theory in a region of weak gradients using this assumption.
In normalised quantities, \eqref{radialForceBalance} gives a relation between $\bar{u}$ and the normalised pressure gradient, 
\begin{equation}\label{Holy u}
    \pdv{}{\bar{\psi}}\ln \bar{p}=-2\frac{\bar{u}}{\bar{T}}.
\end{equation}
Using the input profiles of density and temperature, \eqref{Holy u} determines $\bar{u}$ which can be used together with the input profile for $\bar{V}$ in the normalised quasineutrality equation \eqref{HolyPhi} to determine the poloidal variation of the electric potential. This enables us to calculate the neoclassical ion particle and energy fluxes from \eqref{HolyGamma} and \eqref{HolyQ}. The last equation is \eqref{Holygamma} and gives the parallel momentum input $\bar \gamma$ to maintain the particle flux. \par 
 
 The second approach is to assume vanishing parallel momentum source, $\bar\gamma=0$. \cite{Trinczek2023} showed that without any input of parallel momentum, \eqref{Holygamma} leads to
\begin{equation}\label{NA}
\bar \Gamma\simeq0,    
\end{equation} 
to lowest order in $\sqrt{m_e/m}$. We call the resulting particle transport neoclassically ambipolar \citep{trinczek2025a} because the neoclassical ion particle transport in this scenario is negligible to lowest order, as in the core, where intrinsic ambipolarity enforces balance between the neoclassical ion and electron particle fluxes. The implications of $\bar\gamma=0$ for the particle and momentum transport are discussed in detail also in \cite{trinczek2026}. The expressions for $\bar{\phi}_c$ and $\bar{\phi_s}$ are obtained as functions of $\bar{u}$ from \eqref{HolyPhi} and are substituted into \eqref{HolyGamma} and the resulting nonlinear equation is solved for $\bar{u}$. Once $\bar{u}$ is determined, it is substituted back into \eqref{HolyPhi} and finally \eqref{HolyQ} to calculate the energy flux.\par

\begin{figure}
    \centering
    \includegraphics[width=0.48\linewidth]{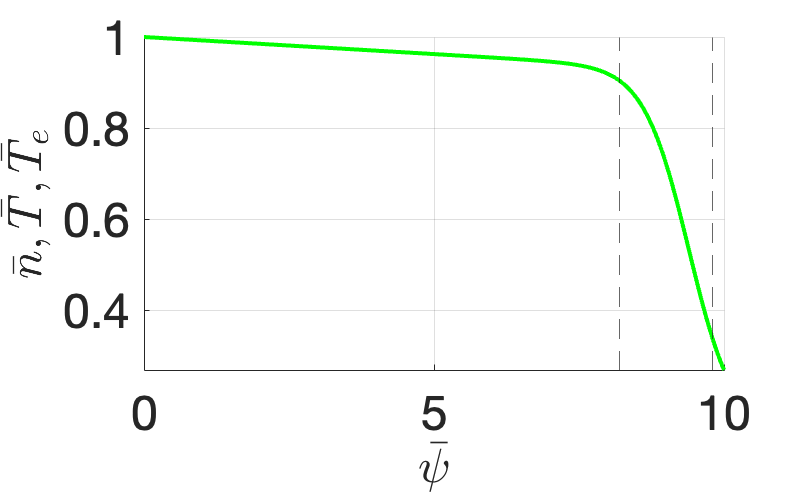}
    \includegraphics[width=0.48\linewidth]{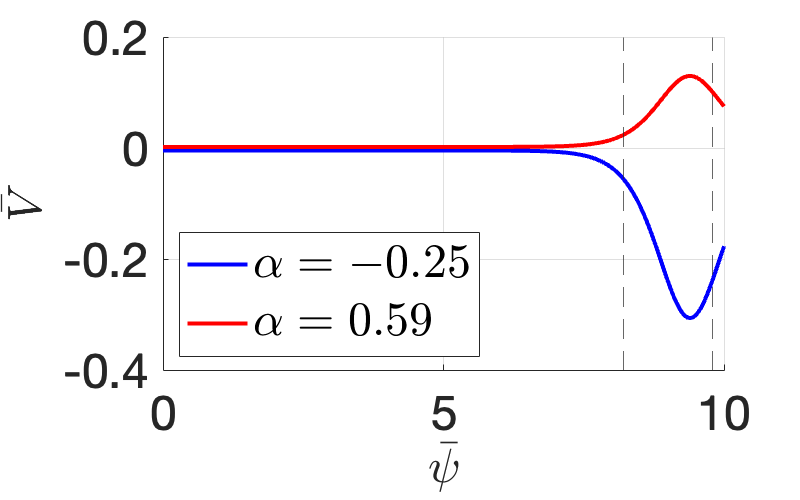}
    \caption{Input profiles for normalised ion and electron temperature and density from \eqref{input}. The strong gradient region is indicated by vertical dashed lines. We compare two different input profiles for the mean parallel flow of the form \eqref{Vinput}.}
    \label{fig:Profiles}
\end{figure}

To study the new equations for large gradient neoclassical theory, we consider the example density and ion and electron temperature
\begin{equation}\label{input}
    \bar{n}=\bar{T}=\bar{T}_e= 0.6035+0.3965\tanh[-1.2929(\bar{\psi}-9.3942)]-0.0075\bar{\psi}.
\end{equation}
The profiles are shown in figure \ref{fig:Profiles}.

\par Instead of solving for $\bar{V}$, we use reasonable assumptions of what the mean parallel flow might be. We want to emphasise that the mean parallel flow in this approach is an input. It cannot be easily determined in strong gradient regions due to the nonlinear character of the transport equations, as explained in \cite{trinczek2025a} and \cite{trinczek2026}. For the mean parallel flow, we choose
\begin{equation}\label{Vinput}
    \bar{V}=\alpha\pdv{\bar{T}}{\bar{\psi}},
\end{equation}
where $\alpha$ is a parameter.
We compare two different choices for the mean parallel flow for $\alpha=-0.25$ and $\alpha=0.59$. These choices for the mean parallel flow are arbitrary but motivated by the weak gradient result of neoclassical transport in the plateau regime (see Appendix \ref{sec: weak grad}) and the banana regime. If we assume \eqref{radialForceBalance}, the weak gradient mean parallel flow in \eqref{Vwg} reduces to \eqref{Vinput} with $\alpha=-0.25$ in the plateau regime. Similarly, the weak gradient mean parallel flow in the banana regime with \eqref{radialForceBalance} reduces to \eqref{Vinput} with $\alpha=0.59$. We plot $\bar V$ for $\alpha=-0.25$ and 0.59 in figure \ref{fig:Profiles}. As we will demonstrate in this section, the choice of $\bar{V}$ is crucial because $\bar V$ determines to great extent, whether neoclassical transport in strong gradient regions deviates significantly from weak gradient results.\par

We will analyse radial electric field, fluxes, poloidal variation and bootstrap current for both choices of $\alpha$ in \eqref{Vinput}. First, we follow the approach of radial force balance and then compare the results to those we get from applying neoclassical ambipolarity.

\subsection{Radial force balance}\label{sec: Force Balance}
\begin{figure}
    \centering
    \includegraphics[width=0.5\linewidth]{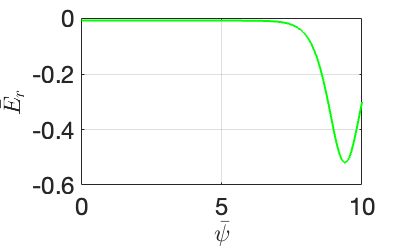}
    \caption{The radial electric field as determined by radial force balance \eqref{radialForceBalance}.}
    \label{fig:ErFB}
\end{figure}

Radial force balance as defined in \eqref{radialForceBalance} gives the radial electric field directly from the pressure gradient. The radial electric field for the input profiles in figure \ref{fig:Profiles} is shown in figure \ref{fig:ErFB}. We observe a deep radial electric field well, as observed in pedestals \citep{mcdermott2009, Viezzer2013}. Radial force balance is independent of the mean parallel flow, so the radial electric field is the same for any choice of $\alpha$ in \eqref{Vinput}.

 \par
\begin{figure}
    \centering
    \includegraphics[width=0.48\linewidth]{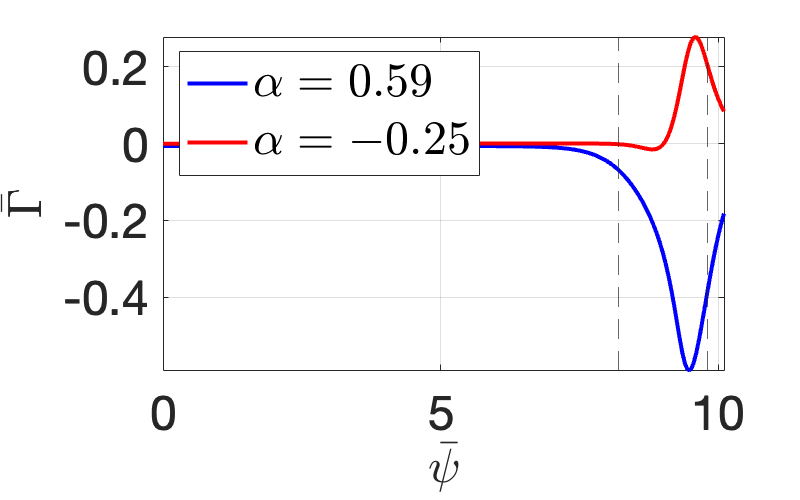}
    \includegraphics[width=0.48\linewidth]{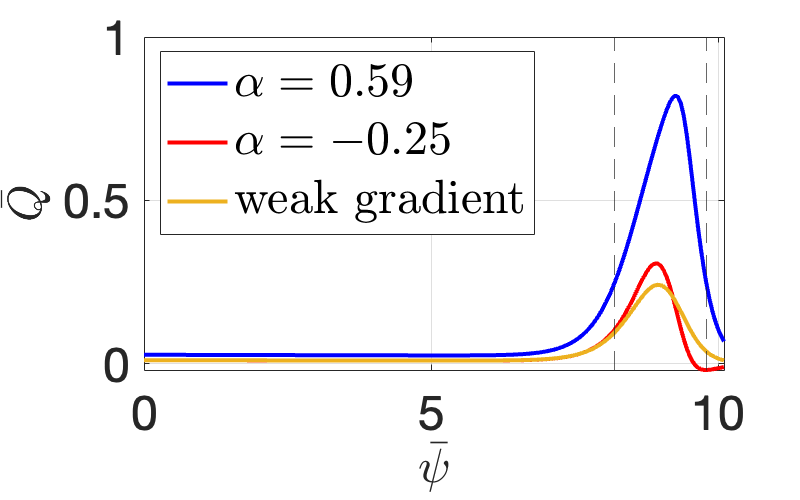}
    \caption{The ion neoclassical particle flux is mostly positive for $\alpha=-0.25$ and negative for $\alpha=0.59$. The ion neoclassical energy flux exceeds the weak gradient energy flux in the strong gradient region.}
    \label{fig:FluxesFB}
\end{figure}

The ion neoclassical particle and energy fluxes are shown in figure \ref{fig:FluxesFB}. The ion neoclassical particle flux is mostly positive for $\alpha=-0.25$ and negative for $\alpha=0.59$. Radial force balance thus implies that turbulence must exist in the system such that ambipolarity is achieved through a balance between the neoclassical particle flux (which we calculate) and the turbulent particle flux (not calculated in this paper) -- see \cite{trinczek2026} for a more extensive discussion.\par 
We can compare the energy flux for both values of $\alpha$ with the weak gradient result \eqref{Qwg}. The weak gradient energy flux \eqref{Qwg} does not explicitly depend on the mean parallel flow because the flow is known in the weak gradient limit, see \eqref{Vwg}. This is not the case in the strong gradient neoclassical expression \eqref{HolyQ}, where the dependence on $V_\parallel$ is kept explicitly. The strong gradient neoclassical energy flux exceeds the weak gradient theory result for both values of $\alpha$. The maximum energy flux exceeds the maximum energy flux according to weak gradient theory for $\alpha=-0.25$ by $\bar{Q}^{FB}_{max}/\bar{Q}^{wg}_{max}\simeq 1.27$, and by $\bar{Q}^{FB}_{max}/\bar{Q}^{wg}_{max}=3.41$ for $\alpha=0.59$. This is a significant increase of the neoclassical energy flux when strong gradient effects are kept. Clearly, the energy fluxes within strong gradient neoclassical theory depend strongly on the mean parallel flow. It is also worth noting that \cite{Seol2012} claimed that strong gradient effects will always reduce the energy flux in comparison to weak gradient neoclassical theory. Our example case is in clear contradiction to this result. Further discussion of this discrepancy is provided in Appendix \ref{sec: Comparison Peter/Shaing}.\par
\begin{figure}
    \centering
    \includegraphics[width=0.48\linewidth]{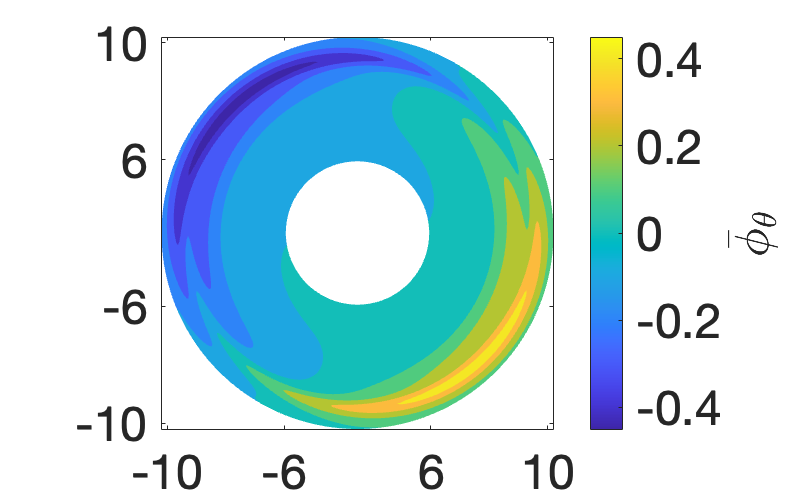}
    \includegraphics[width=0.48\linewidth]{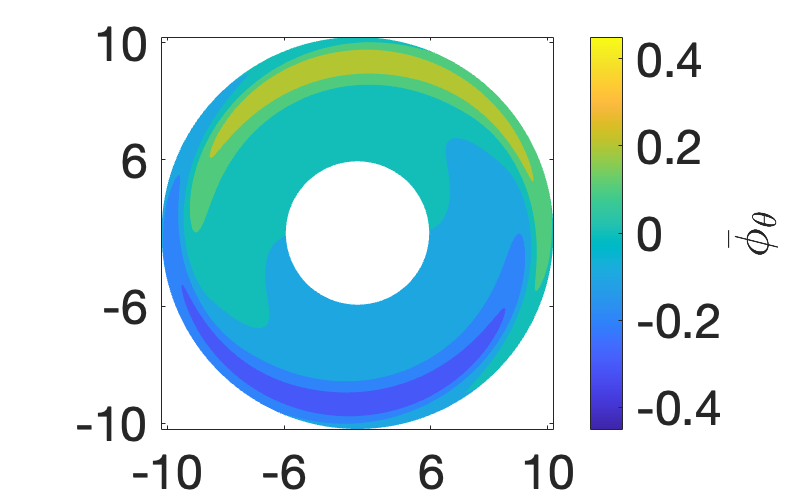}
       \includegraphics[width=0.48\linewidth]{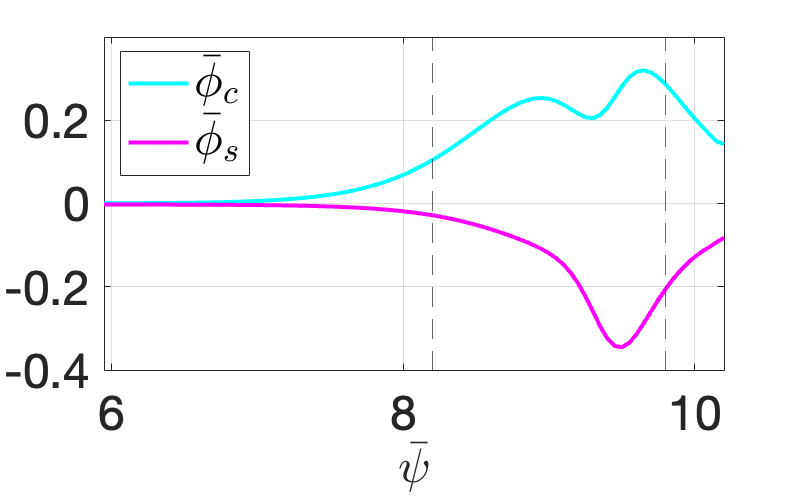}
    \includegraphics[width=0.48\linewidth]{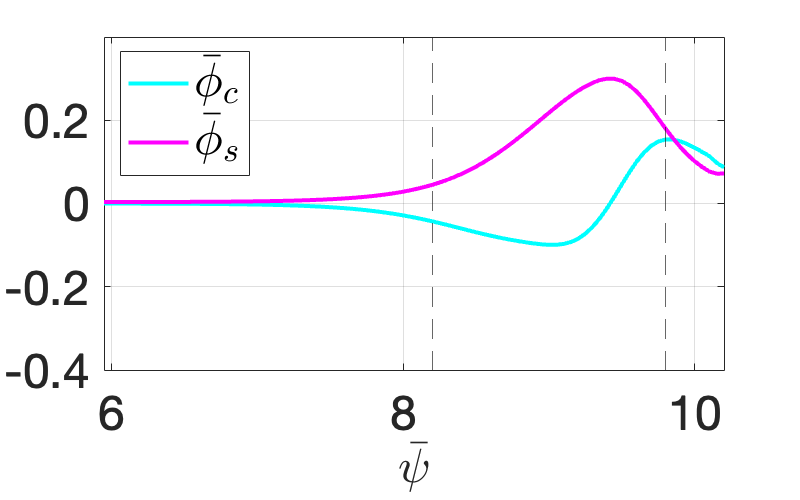}
    \caption{On the left is the poloidal variation of the electric potential $\bar{\phi}_\theta=\bar{\phi}_c\cos\theta+\bar{\phi}_s\sin\theta$ for $\alpha=0.59$, and on the right for $\alpha=-0.25$. Both cases show in-out and up-down asymmetry. The corresponding amplitudes $\bar\phi_c$ and $\bar\phi_s$ are shown as a function of $\bar\psi$ below the respective two dimensional plots.}
    \label{fig:PoloidalVarFB}
\end{figure}

We can use \eqref{potential assumption plateau}, \eqref{phic plateau} and \eqref{phis plateau} to obtain the poloidal variation of the electric potential for force balance for the two choices of $\alpha$. In figure \ref{fig:PoloidalVarFB}, we show $\bar{\phi}$ as a function of $\bar{\psi}$ and $\theta$ for $\bar{\psi}\in [6,10]$ as well as the respective amplitudes $\bar\phi_c$ and $\bar\phi_s$. Note that figure \ref{fig:PoloidalVarFB} is not true to scale in radius -- the edge is enlarged relative to the rest of the poloidal plane. There is stronger in-out asymmetry for $\alpha=-0.25$ and stronger up-down asymmetry for $\alpha=0.59$.\par
\begin{figure}
    \centering
    \includegraphics[width=0.48\linewidth]{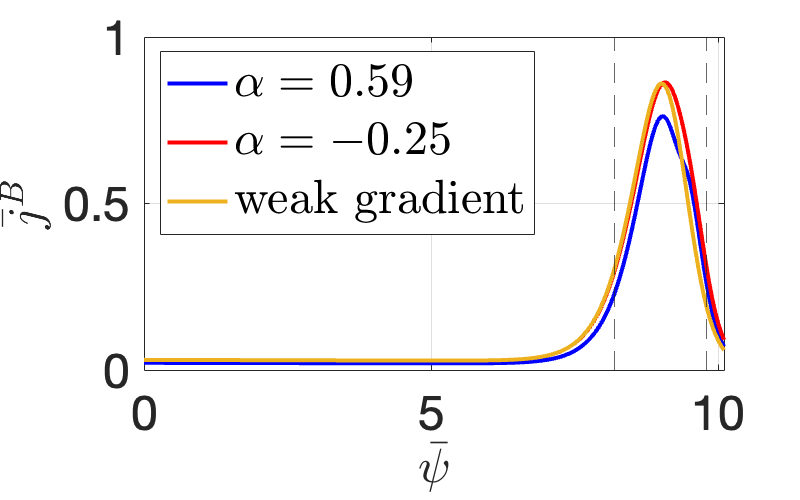}
    \includegraphics[width=0.48\linewidth]{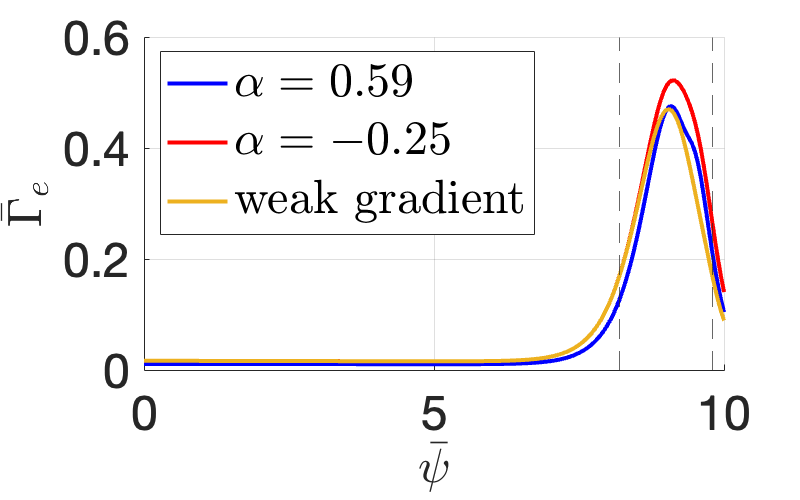}
    \caption{The bootstrap current is almost identical to weak gradient neoclassical theory for $\alpha=-0.25$ but smaller than weak gradient neoclassical predictions for $\alpha=0.59$. The electron neoclassical particle flux is closer to the weak gradient limit for $\alpha=0.59$ and larger for $\alpha=-0.25$.}
    \label{fig:BootstrapFB}
\end{figure}

The bootstrap current is shown in figure \ref{fig:BootstrapFB}. Again, we can compare our modification to the limit of weak gradient neoclassical theory \eqref{jwg}. For $\alpha=-0.25$, the changes are not significant. However, for $\alpha=0.59$, the maximum bootstrap current calculated with strong gradient neoclassical theory is smaller than the peak bootstrap current predicted by weak gradient neoclassical theory by $\bar{j}^{B,NA}_{max}/\bar{j}^{B,wg}_{max}=0.89$. In this case, keeping strong gradient effects results in a lower bootstrap current than weak gradient neoclassical theory predicts.\par 
The electron neoclassical particle flux, which is also plotted in figure \ref{fig:BootstrapFB}, shows little to no deviation from weak gradient neoclassical theory for both $\alpha=-0.25$ and $\alpha=0.59$. The neoclassical electron particle flux is just one contribution to the total electron flux and cannot balance the neoclassical ion particle flux by itself due to the smallness of $\Gamma_e$ in the mass ratio. The turbulent electron particle flux has to be of the same size as the ion neoclassical particle flux to satisfy ambipolarity.

\subsection{Neoclassical Ambipolarity} \label{sec: NA}
\begin{figure}
    \centering
    \includegraphics[width=0.48\linewidth]{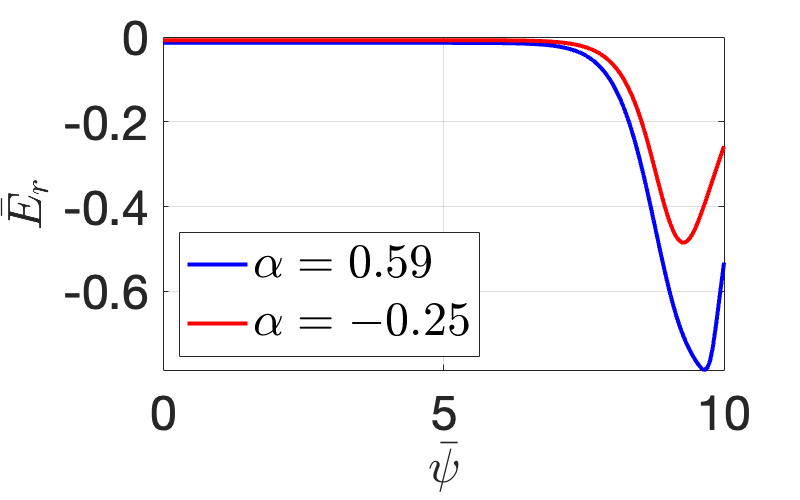}
    \caption{The radial electric field is shown as determined by neoclassical ambipolarity \eqref{NA} for $\alpha=0.59$ and $\alpha=-0.25$.}
    \label{fig:ErNA}
\end{figure}

In the absence of a parallel momentum source, neoclassical ambipolarity \eqref{NA} can be used to determine the radial electric field by solving $\bar \Gamma=0$, with $\bar \Gamma$ given by \eqref{HolyGamma}, for $\bar u$ for a specified $\bar V$. This approach was not possible in Sec. \ref{sec: Force Balance} because $\bar \Gamma$ was not known and hence force balance was used to determine the radial electric field. Solving \eqref{HolyGamma} for $\bar\Gamma=0$ gives the radial electric field profiles for $\alpha=-0.25$ and $\alpha=0.59$ in figure \ref{fig:ErNA}. Here, the radial electric field depends on the choice of the mean parallel flow and we get a larger radial electric field for $\alpha=0.59$. We can compare these profiles with radial force balance in figure \ref{fig:ErFB} and find that neoclassical ambipolarity predicts a weaker radial electric field for $\alpha=-0.25$ and a stronger one for $\alpha=0.59$.

\begin{figure}
    \centering
    \includegraphics[width=0.48\linewidth]{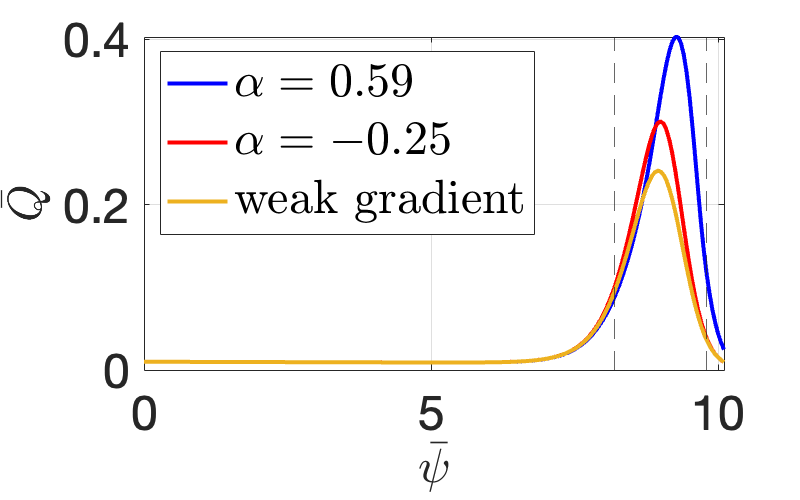}
    \caption{The ion energy flux exceeds the weak gradient neoclassical prediction for $\alpha=0.59$ and for $\alpha=-0.25$.}
    \label{fig:FluxesNA}
\end{figure}

By definition, the ion neoclassical particle flux vanishes to lowest order for neoclassical ambipolarity. The ion neoclassical energy flux is shown in figure \ref{fig:FluxesNA}. Again, the energy flux with strong gradient effects exceeds the weak gradient prediction for both choices of the mean parallel flow. For $\alpha=0.59$, the energy flux exceeds the weak gradient theory prediction by $\bar{Q}^{NA}_{max}/\bar{Q}^{wg}_{max}=1.61$. For $\alpha=-0.25$, the energy flux is very similar to the one we determined using force balance and $\bar{Q}^{NA}_{max}/\bar{Q}^{wg}_{max}\simeq 1.25$. This shows that the mean parallel flow has a strong impact on the deviation of fluxes from weak gradient neoclassical theory in the case of neoclassical ambipolarity as well. 

\begin{figure}
    \centering
    \includegraphics[width=0.48\linewidth]{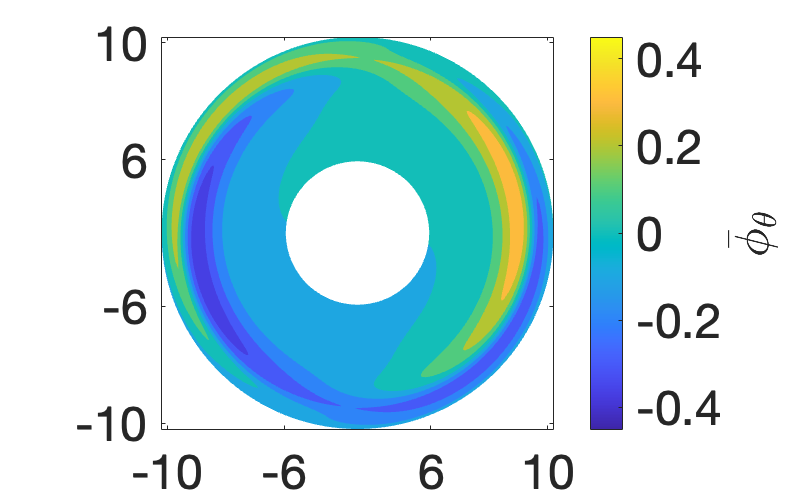}
    \includegraphics[width=0.48\linewidth]{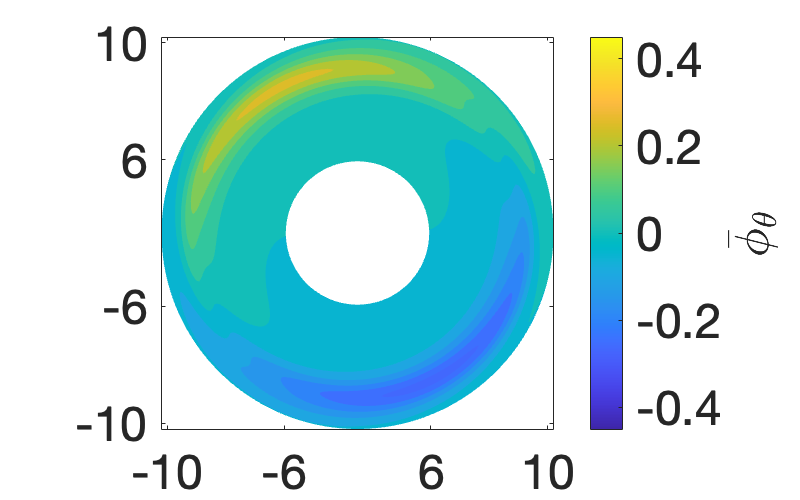}
     \includegraphics[width=0.48\linewidth]{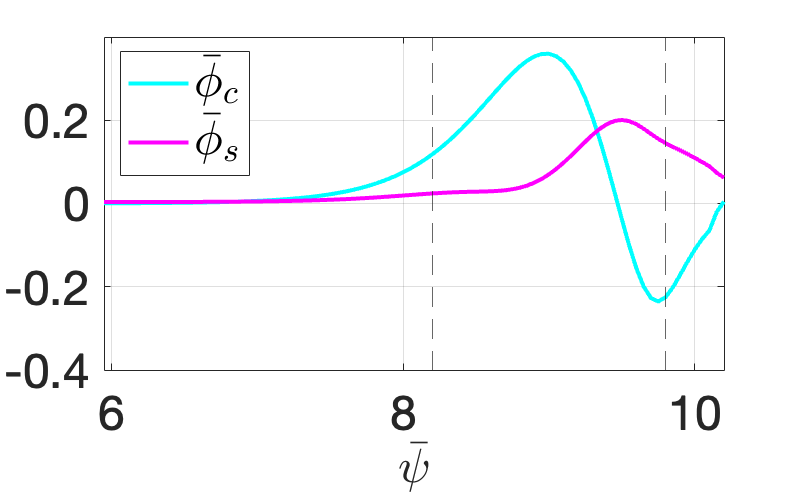}
    \includegraphics[width=0.48\linewidth]{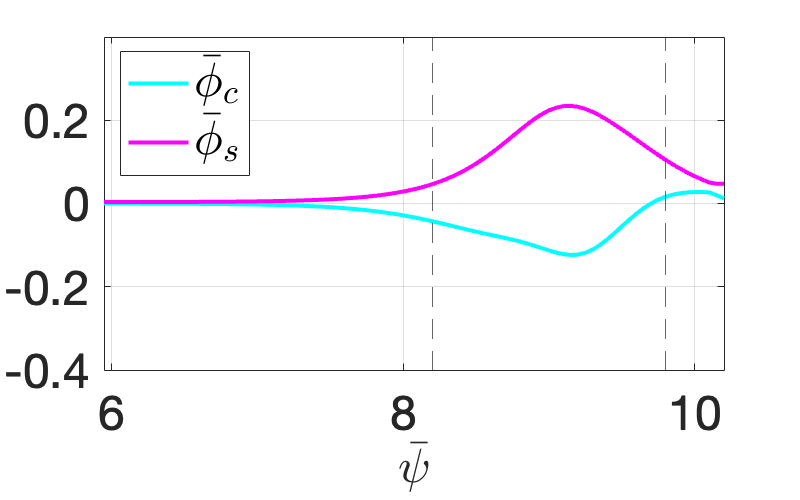}
    \caption{On the left is the poloidal variation of the electric potential $\bar{\phi}_\theta=\bar{\phi}_c\cos\theta+\bar{\phi}_s\sin\theta$ for $\alpha=0.59$. The amplitudes $\bar\phi_c$ and $\bar\phi_s$ are given as functions of $\bar\psi$ in the lower figures. On the right is the poloidal variation for $\alpha=-0.25$ which shows stronger up-down asymmetry and hence a larger absolute value of $\bar\phi_s$.}
    \label{fig:PoloidalVarNA}
\end{figure}

The poloidal variation of the electric potential for neoclassical ambipolarity is shown in figure \ref{fig:PoloidalVarNA}. For $\alpha=0.59$, the variation is more in-out asymmetric whereas we observe stronger up-down asymmetry for $\alpha=-0.25$. Overall, the peak amplitudes of the variation is weaker than with radial force balance. Noticeably, the amplitude for up-down asymmetry is positive for neoclassical ambipolarity and negative for radial force balance for $\alpha=0.59$. For $\alpha=-0.25$, the poloidal variation is qualitatively fairly similar for radial force balance and neoclassical ambipolarity.

\begin{figure}
    \centering
    \includegraphics[width=0.48\linewidth]{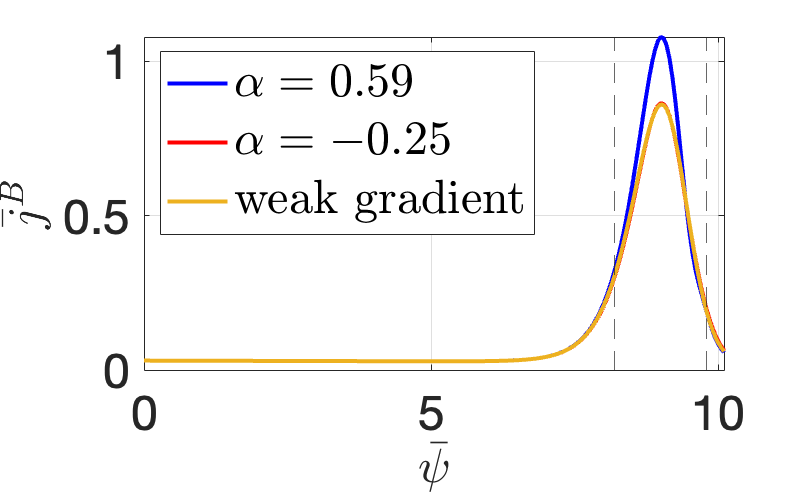}
    \includegraphics[width=0.48\linewidth]{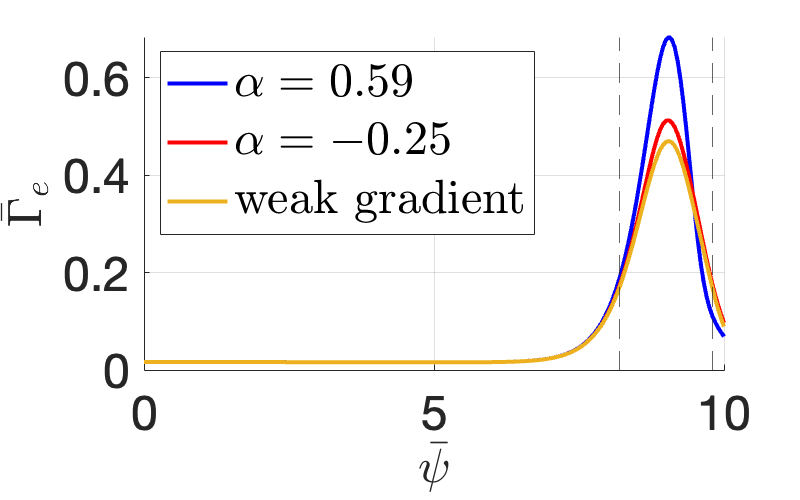}
    \caption{The bootstrap current exceeds weak gradient neoclassical predictions for $\alpha=0.59$ and is indistinguishable from weak gradient theory for $\alpha=-0.25$. The neoclassical electron flux exceeds weak gradient results in both cases, but more so for $\alpha=0.59$.}
    \label{fig:BootstrapNA}
\end{figure}

The bootstrap current is almost identical to weak-gradient neoclassical theory for $\alpha=-0.25$, as can be seen in figure \ref{fig:BootstrapNA}. This is similar to the case where force balance was used. However, for $\alpha=0.59$ we observe an increased bootstrap current by $\bar{j}^{B,NA}_{max}/\bar{j}^{B,wg}_{max}=1.25$. This is the opposite to what we observed in radial force balance, where a reduced bootstrap current was found for $\alpha=0.59$.\par
The electron neoclassical particle flux strongly exceeds the weak gradient prediction for $\alpha=0.59$.  

\section{Conclusion}\label{sec: Conclusion}
Neoclassical transport theory has been extended into regions of strong gradients where the length scale of density, potential and temperature gradients is on the order of the ion poloidal gyroradius. This approach separates the $\rho_p$ and $\rho$ length scales by employing a large aspect ratio expansion, which makes an analytical treatment possible. Poloidal variation has been kept and the mean parallel flow is allowed to be of the order of the thermal speed. We find that strong gradient effects can enhance or reduce neoclassical transport in comparison to weak gradient predictions depending on the input profiles.\par

In section \ref{sec: General equations}, we derive limits of the drift kinetic equation that give the ion and electron distribution functions for $\nu_\ast\sim 1$. General expressions for the neoclassical transport of ions and electrons as well as the bootstrap current are derived from moments of the drift kinetic equation. The transport relations are valid in both the banana and the plateau regime. The general discussion was followed by the derivation of the distribution function and transport relations in the plateau regime in section \ref{sec: Plateau}. The main differences with weak gradient neoclassical theory are the modification due to poloidal variation and the dependence on the mean parallel flow.

We conduct a study of realistic input profiles for density, temperature and parallel flow to understand the impact of strong gradient modifications to weak gradient neoclassical theory. We introduce two different approaches to determine the radial electric field. The first approach uses radial force balance to calculate $E_r$ from the pressure gradient. The second approach is to impose neoclassical ambipolarity, where $E_r$ is determined from vanishing neoclassical ion particle flux. We find that, depending on the mean parallel flow, deviations from weak gradient neoclassical theory can be significant, like in figure \ref{fig:FluxesFB} where the energy flux is increased by roughly a factor 3.41, or negligible, like in figure \ref{fig:BootstrapNA} where the bootstrap current remained unchanged (for $\alpha=-0.25$). We see that the behavior of the mean parallel flow and the closure via force balance or neoclassical ambipolarity has not only a quantitative but a qualitative impact on transport predictions. The poloidal variation, for example, can change from primarily in-out to primarily up-down asymmetric like in figure \ref{fig:PoloidalVarNA}. With these few examples, we cannot draw a universal conclusion on how strongly strong gradient effects modify neoclassical transport, but we demonstrated that strong enhancement or decrease can occur for realistic pedestal profiles.\par

One of the main differences between the plateau and the banana regime in these equations is the poloidal variation of the electric potential. In the plateau regime, a mix of in-out and up-down symmetry is present whereas only in-out asymmetry is possible in the banana regime.
It is apparent from \eqref{gammac1} that the neoclassical ion particle flux has the same dependence on the parallel momentum input as in the banana regime. Parallel momentum input and radial particle flux are strongly connected. Turbulence or impurities can affect the parallel momentum equation and enable a dominant neoclassical ion particle flux. \cite{trinczek2026} provide an extensive discussion of the ion neoclassical particle flux and its relation to parallel momentum sources.\par

 The transport relations agree nicely with weak gradient neoclassical theory in the limit of weak gradients, as shown in Appendix \ref{sec: weak grad}. We compare our approach with previous work that allowed stronger gradients for neoclassical theory by \cite{Pusztai2010} and \cite{Seol2012} in Appendix \ref{sec: Comparison Peter/Shaing}, and find that our theory supersedes these approaches. In previous work, poloidal variation was neglected and weak mean parallel flow and neoclassical ambipolarity were assumed. Even in the limits where these works were derived, we find discrepancies. In the limit of weak temperature gradient, weak mean parallel flow, neoclassical ambipolarity and zero poloidal variation, we find agreement with a version of \cite{Pusztai2010} before an incorrect corrigendum. In the limit of weak mean parallel flow and neoclassical ambipolarity, we disagree with \cite{Seol2012} due to a problem in the moment approach they used, which we explain in detail in Appendix \ref{sec: Comparison Peter/Shaing}. We conclude that our model is the most complete and correct extension of neoclassical theory into regions of strong gradients for large aspect ratio tokamaks in the plateau regime.

 \section*{Funding}
This work was supported by the U.S. Department of Energy  Laboratory Directed Research and Development program at the Princeton Plasma Physics Laboratory (S.T. and F.I.P., contract number DE-AC02-09CH11466 and P.J.C., contract number DE-FG02-91ER-54109). The United States Government retains a non-exclusive, paid-up, irrevocable, world-wide license to publish or reproduce the published form of this manuscript, or allow others to do so, for United States Government purposes. 

\section*{Declaration of Interests}
The authors report no conflict of interest.

\appendix
\section{The distribution function $g^{t,bp}_1$}\label{sec: g1}
We need to prove that the $\theta$-dependent part of $g^{t,bp}_1$ decays as $w\rightarrow\pm \infty$. The lowest order drift kinetic equation \eqref{drift g conservative trapped} which includes $g^{t,bp}_1$ reads
\begin{multline}\label{DKE g1}
    \pdv{}{\theta}\left(\frac{w}{qR} g_1^{t,bp}\right)+\pdv{u}{\theta}\pdv{}{w}\left(\frac{w}{qR}g_0^{t,bp}\right)-S\pdv{}{w}\left[\mathcal{P}(\theta)g_1^{t,bp}\right] +\pdv{}{\psi}\left[\frac{I}{\Omega}\mathcal{P}(\theta)g^{t,bp}_0\right]\\
    =\pdv{}{w}\left[\boldsymbol{\hat{b}}\bcdot\bm{M}^{t,bp}\right]+\pdv{}{\mu}\left[f_M 2\mu\mathsf{M}_\perp\pdv{(g^{t,bp}_0/f_M)}{w}\right],
\end{multline}
where we introduced the perpendicular component of $\mathsfbi{M}$
\begin{equation}
    \mathsf{M}_\perp\equiv \frac{\bm{v}_\perp}{\abs{\bm{v}_\perp}^2}\bcdot \mathsfbi{M}\bcdot \bm{\hat{b}}\simeq (-u-V_\parallel)\left(\frac{\nu_\parallel}{2}-\frac{\nu_\perp}{4}\right).
\end{equation}
For large $w$, $g_0^{t,bp}$ and $\boldsymbol{\hat{b}}\bcdot\bm{M}^{t,bp}$ tend to a constant to match the slowly varying $g^p$ and $\boldsymbol{\hat{b}}\bcdot\bm{M}^{p}$. The derivative $\partial g_1^{t,bp}/\partial w$ also tends to a constant to match $\partial g^p/\partial w$. Then, for large $w$, \eqref{DKE g1} becomes
\begin{equation}\label{large g1}
    \pdv{}{\theta}\left(\frac{w}{qR} g_1^{t,bp}\right)
    \simeq-\pdv{u}{\theta}\frac{g_0^{t,bp}(w\rightarrow\pm\infty)}{qR}-\pdv{}{\psi}\left[\frac{I}{\Omega}\mathcal{P}(\theta)g^{t,bp}_0(w\rightarrow \pm\infty)\right].
\end{equation}
The lowest order distribution function $g^{t,bp}_0$ has a piece which is independent of $\theta$ and constant at $w\rightarrow \pm\infty$ such that $g^{t,bp}_0(w\rightarrow\infty)-g^{t,bp}_0(w\rightarrow-\infty)=\Delta g^p$. The piece that depends on $\theta$ decays as $1/w$, as explained in \eqref{big w limit}. Thus, the right hand side of \eqref{large g1} is independent of $w$ and we find that the $\theta$-dependent part of $g_1^{t,bp}\sim1/w$ for large $w$.

\section{Plateau regime distribution function}\label{sec: g0}
\subsection{The distribution function $g^l_0$}
The solution to \eqref{Jump Condition Plateau} is the lowest order distribution function $g^l_0$ in the plateau regime. First, we rewrite \eqref{Jump Condition Plateau} in terms of the normalized variable $\xi=w/v_{ref}$ with $v_{ref}=(qR\mathsf{M}_\parallel)^{1/3}$ to find
\begin{equation}\label{normalised jump cond}
    \pdv{}{\theta}\left(\xi g^l_0\right)-\pdv[2]{g^l_0}{\xi}=\frac{I}{\Omega}\left(A_S\sin\theta+A_C\cos\theta\right)\frac{r}{R}\frac{1}{qR}\mathcal{D}f_M(w=0).
\end{equation}
We employ the ansatz 
\begin{equation}\label{gl0 ansatz}
    g^l_0=a(w) \exp(i\theta)+b(w)\exp(-i\theta)
\end{equation}
and separate terms proportional to $\exp(i\theta)$ and $\exp(-i\theta)$ in \eqref{normalised jump cond}. This yields two differential equations for the amplitudes $a(w)$ and $b(w)$,
\begin{equation}\label{eq a}
    i\xi a-\pdv[2]{a}{\xi}=\frac{I}{\Omega}\frac{A_C-iA_S}{2}\frac{r}{R}\frac{1}{v_{ref}}\mathcal{D}f_M(w=0)
\end{equation}
and
\begin{equation}\label{eq b}
    -i\xi b-\pdv[2]{b}{\xi}=\frac{I}{\Omega}\frac{A_C+iA_S}{2}\frac{r}{R}\frac{1}{v_{ref}}\mathcal{D}f_M(w=0).
\end{equation}
Equations of the form
\begin{equation}
    \pm i\xi h-\pdv[2]{h}{\xi}=1
\end{equation}
have been solved by \cite{su1968}, who found the solutions
\begin{equation}\label{SuObermann}
h=\int_0^\infty\mathrm{d}p\:\exp\left(-\frac{p^3}{3}\mp i\xi p\right).    
\end{equation}
This solution can be checked by substitution
\begin{multline}
     \pm i\xi h-\pdv[2]{h}{\xi}=\int_0^\infty\mathrm{d}p\:\exp\left(-\frac{p^3}{3}\mp i\xi p\right)(p^2\mp i\xi)\\
     =-\int_0^\infty\mathrm{d}p\: \pdv{}{p}\left[\exp\left(-\frac{p^3}{3}\mp i\xi p\right)\right]=1.
\end{multline}
We can use \eqref{SuObermann} and find for $g^l_0$ in \eqref{gl0 ansatz}
\begin{multline}
    g^l_0=\frac{I}{\Omega}\frac{r}{R}\frac{1}{v_{ref}}\mathcal{D}f_M(w=0)\bigg\lbrace\int_0^\infty\mathrm{d}p\:\frac{A_C-iA_S}{2}\exp\left[-\frac{p^3}{3}+i(\theta-p\xi)\right]\\
    +\int_0^\infty\mathrm{d}p\:\frac{A_C+iA_S}{2}\exp\left[-\frac{p^3}{3}-i(\theta-p\xi)\right]\bigg\rbrace,
\end{multline}
which can be rearranged to the form in \eqref{gl}.\par
For completeness, we also show how one can solve the differential equations \eqref{eq a} and \eqref{eq b} for $a$ and $b$ via Fourier transform from $\xi$ to $p$. When we multiply \eqref{eq a} and \eqref{eq b} by $\exp(ip\xi)/2\pi$ and integrate over $\xi$, we find
\begin{equation}
    \pdv{a_p}{p}+p^2 a_p=\delta(p)\frac{I}{\Omega}\frac{A_C-iA_S}{2}\frac{r}{R}\frac{1}{v_{ref}}\mathcal{D} f_M(w=0)
\end{equation}
and
\begin{equation}
    -\pdv{b_p}{p}+p^2 b_p=\delta(p)\frac{I}{\Omega}\frac{A_C+iA_S}{2}\frac{r}{R}\frac{1}{v_{ref}}\mathcal{D} f_M(w=0).
\end{equation}
Here, $a_p$ and $b_p$ are the Fourier transforms of $a$ and $b$ and $\delta(p)$ is the Dirac delta function. The solutions to these ordinary differential equations that vanish for $p\rightarrow \pm \infty$ are
\begin{equation}
    a_p(p)=H(p)\exp\left(-\frac{p^3}{3}\right)\frac{I}{\Omega}\frac{A_C-iA_S}{2}\frac{r}{R}\frac{1}{v_{ref}}\mathcal{D}f_M(w=0)
\end{equation}
and
\begin{equation}
    b_p(p)=H(-p)\exp\left(\frac{p^3}{3}\right)\frac{I}{\Omega}\frac{A_C+iA_S}{2}\frac{r}{R}\frac{1}{v_{ref}}\mathcal{D}f_M(w=0).
\end{equation}
Here, $H(p)$ is the Heaviside step function. The inverse Fourier transform gives the result
\begin{equation}
    g^l_0=\int_{-\infty}^\infty\mathrm{d}p\:a_p(p)\exp(i\theta-ip\xi)+\int_{-\infty}^\infty\mathrm{d}p\:b_p(p)\exp(-i\theta-ip\xi),
\end{equation}
which can also be rearranged to the form in \eqref{gl}.\par
The distribution function $g^l_0$ in \eqref{gl} matches with \eqref{theta dependent gp} for $\xi\rightarrow\infty$, as can be seen by changing the integration variable $p$ to $t=p\xi$ in \eqref{Fs} and \eqref{Fc}, finding
\begin{multline}
F_s(\xi,\theta)=\int_0^\infty\mathrm{d}p \:\exp\left(-\frac{p^3}{3}\right)\sin(\theta-p\xi)\\
=\int_0^\infty\frac{\mathrm{d}t}{\xi} \:\exp\left(-\frac{t^3}{3\xi^3}\right)\sin(\theta-t)\simeq-\frac{\cos\theta}{\xi}
\end{multline}
and
\begin{multline}
F_c(\xi,\theta)=\int_0^\infty\mathrm{d}p \:\exp\left(-\frac{p^3}{3}\right)\cos(\theta-p\xi)\\
=\int_0^\infty\frac{\mathrm{d}t}{\xi} \:\exp\left(-\frac{t^3}{3\xi^3}\right)\cos(\theta-t)\simeq\frac{\sin\theta}{\xi}
\end{multline}
for $\xi\rightarrow\infty$. Thus, the large $w$ limit of \eqref{gl} is
\begin{equation}
    g^l_{0}\simeq-\left(A_S\cos\theta-A_C\sin\theta\right)\frac{I}{\Omega} \frac{\mathcal{D}}{w}\frac{r}{R} f_M(w=0),
\end{equation}
whereas the limit of small $w$  of the $\theta$-dependent freely passing distribution function in \eqref{theta dependent gp} is
\begin{equation}\label{theta gp limit}
        g^p-\langle g^p\rangle_\tau        \simeq-\frac{I}{\Omega}\frac{r}{R}\frac{A_S \cos{\theta}-A_C\sin{\theta}}{w}\mathcal{D}f_M(w=0).
\end{equation}
We find that the large $w$ limit of $g^l_0$ matches with the small $w$ limit of \eqref{gl}. 
\subsection{Integrals over $g^l_0$}
We need the integral 
\begin{equation}\label{Sin Int}
    \int\mathrm{d}w\:g^l_0\propto \int_{-\infty}^\infty\mathrm{d}\xi\: \int_0^\infty \mathrm{d}p\: \exp\left(-\frac{p^3}{3}\right)\left[A_S\sin(\theta-p\xi)+A_C\cos(\theta-p\xi)\right]
\end{equation}
to calculate the particle transport in \eqref{Gamma g0}. Using trigonomic identities and the fact that the integral over $\sin(p\xi)$ vanishes because it is odd in $\xi$, we find
\begin{equation}
    \int\mathrm{d}w\:g^l_0\propto \int_{-\infty}^\infty\mathrm{d}\xi\: \int_0^\infty \mathrm{d}p\: \exp\left(-\frac{p^3}{3}\right)\left(A_S\sin\theta+A_C\cos\theta\right)\cos(p\xi).
\end{equation}
The remaining integral is an integral over $\exp(-p^3/3)\cos(p\xi)$ and can be solved by integrating by parts in $p$ and then taking the integral over $\xi$,
\begin{multline}
    \int_{-\infty}^\infty\mathrm{d}\xi\: \int_0^\infty \mathrm{d}p\: \exp\left(-\frac{p^3}{3}\right)\cos(p\xi)=\int_{-\infty}^\infty\mathrm{d}\xi\: \int_0^\infty \mathrm{d}p\: p^2 \exp\left(-\frac{p^3}{3}\right)\frac{\sin(p\xi)}{\xi}\\
    =\pi \int_0^\infty\mathrm{d}p\:p^2\exp\left(-\frac{p^3}{3}\right)=\pi.
\end{multline}
We arrive at the expression \eqref{int gl}.

\section{Weak gradient limit} \label{sec: weak grad}
In the limit of weak gradients and low flow, $u$, $V_\parallel$ and the pressure and temperature gradients are small. The ion neoclassical particle flux equation \eqref{Gamma} reduces to
\begin{equation}
    \Gamma^{wg}=-\frac{\sqrt{\pi}}{4}\frac{n}{qR}\frac{I^2}{\Omega^2}\left(\frac{r}{R}\right)^2\left(\frac{2T}{m}\right)^{3/2}\left[\pdv{}{\psi}\ln p+\frac{m(u+V_\parallel)}{T}\frac{\Omega}{I}+\frac{1}{2}\pdv{}{\psi}\ln T\right].
\end{equation}
In the weak gradient limit, $\Gamma^{wg}\simeq0$ because it must balance the electron neoclassical particle flux. Hence, we find
\begin{equation}\label{Gammawg zero}
\pdv{}{\psi}\ln p+\frac{m(u+V_\parallel)}{T}\frac{\Omega}{I}=-\frac{1}{2}\pdv{}{\psi}\ln T.
\end{equation}
We can solve \eqref{Gammawg zero} for the parallel flow
\begin{equation}\label{Vwg}
    V_\parallel^{wg}=-u-\frac{IT}{m\Omega}\left(\pdv{}{\psi}\ln p+\frac{1}{2}\pdv{}{\psi}\ln T\right).
\end{equation}
We substitute \eqref{Gammawg zero} directly into the weak gradient limit of \eqref{Q} to find the weak gradient limit for the energy flux \citep{helander2005}
\begin{equation}\label{Qwg}
    Q^{wg}=-\frac{3\sqrt{\pi}}{4}\frac{nT}{qR}\frac{I^2}{\Omega^2 }\left(\frac{r}{R}\right)^2
       \left(\frac{2T}{m}\right)^{3/2}\pdv{}{\psi}\ln T.
\end{equation}
The weak gradient limit for electron fluxes can be found by taking the limit of small $\phi_c$ and $\phi_s$. Thus, the neoclassical electron particle flux in the weak gradient limit is
\begin{equation}
    \Gamma_e^{wg}= -\frac{\sqrt{\pi}}{4}\frac{n_e}{qR}\frac{I^2}{\Omega_e^2}\left(\frac{r}{R}\right)^2\left(\frac{2T_e}{m_e}\right)^{3/2}\mathcal{D}_{e,1/2}.
\end{equation}
We can use \eqref{Gammawg zero} in $\mathcal{D}_{e,l}$ to write
\begin{equation}\label{De wg}
    \mathcal{D}_{e,l}=\pdv{}{\psi}\ln n_e \left(1+\frac{T}{ZT_e}\right)+(1+l)\pdv{}{\psi}\ln T_e +\frac{3}{2ZT_e}\pdv{T}{\psi}.
\end{equation}
Thus, we recover the weak gradient neoclassical electron particle flux \citep{helander2005}
\begin{multline}
    \Gamma_e^{wg}=-\frac{\sqrt{\pi}}{4}\frac{n_e}{qR}\frac{I^2}{\Omega_e^2}\left(\frac{r}{R}\right)^2\left(\frac{2T_e}{m_e}\right)^{3/2}\bigg[\pdv{}{\psi}\ln n_e \left(1+\frac{T}{ZT_e}\right)\\
    +\frac{3}{2}\pdv{}{\psi}\ln T_e +\frac{3}{2ZT_e}\pdv{T}{\psi}\bigg].
\end{multline}
Likewise, the weak gradient limit of the energy flux \citep{helander2005} is
\begin{multline}
    Q_e^{wg}=-\frac{3\sqrt{\pi}}{4}\frac{n_eT_e}{qR}\frac{I^2}{\Omega_e^2}\left(\frac{r}{R}\right)^2\left(\frac{2T_e}{m_e}\right)^{3/2}\bigg[\pdv{}{\psi}\ln n_e \left(1+\frac{T}{ZT_e}\right)\\
    +\frac{5}{2}\pdv{}{\psi}\ln T_e +\frac{3}{2ZT_e}\pdv{T}{\psi}\bigg].
\end{multline}
For the bootstrap current, all terms in \eqref{bootstrap} proportional to $\phi_c$ and $\phi_s$ can be neglected such that
\begin{multline}
    \langle j_\parallel^B\rangle_\psi^{wg}= -\frac{1}{8}\sqrt{\frac{\pi}{2}}\frac{en_e}{\nu_{ee}qR}\frac{I}{\Omega_e}\left(\frac{r}{R}\right)^2\left(\frac{2T_e}{m_e}\right)^{3/2}\\
    \times \left[2a_0\mathcal{D}_{e,\frac{1}{2}}-a_1\mathcal{D}_{e,\frac{13}{2}}-\sum_{i=2} \frac{a_i\Gammaf(i-\frac{1}{2})}{\sqrt{\pi}\Gammaf(1+i)} \mathcal{D}_{e,\gamma_i} \right],
\end{multline}
where $\mathcal{D}_{e,l}$ is given in \eqref{De wg}. We can simplify this expression if we only keep terms proportional to $a_0$ and $a_1$ and thus neglect smaller corrections in the Spitzer function. We can then write the bootstrap current in the weak gradient limit as
\begin{multline}\label{jwg}
    \langle j_\parallel^B\rangle_\psi^{wg}\simeq -\frac{1}{8}\sqrt{\frac{\pi}{2}}\frac{en_e}{\nu_{ee}qR}\frac{I}{\Omega_e}\left(\frac{r}{R}\right)^2\left(\frac{2T_e}{m_e}\right)^{3/2}(2a_0-a_1)\\
    \times \left[\frac{1}{p_e}\pdv{}{\psi}\left(p_i+p_e\right)+\frac{1}{2ZT_e}\pdv{T}{\psi}+\frac{2a_0-13a_1}{4a_0-2a_1}\pdv{}{\psi}\ln T_e\right],
\end{multline}
which agrees with the weak gradient result \citep{hinton1976, Pusztai2010}.

\section{Comparison to previous work on strong gradient neoclassical transport in the plateau regime}\label{sec: Comparison Peter/Shaing}
\subsection{Comparison with the work of \cite{Pusztai2010}}
A neoclassical treatment of regions with strong radial electric field and density gradient in the plateau regime was presented in a paper by \cite{Pusztai2010} with a summary of results provided in \cite{Catto2011}. In their framework, the poloidal variation of the potential was set to zero, the temperature gradient, the parallel flow and $\partial \ln p/\partial\psi+m\Omega u/I$ were assumed to be small and the ion neoclassical particle flux was set to zero. With these assumptions, \eqref{Gamma} gives
\begin{equation}\label{D1}
0=\frac{m^2u^4}{2T^2}\mathcal{D}_{-3/2}+\frac{mu^2}{T}\mathcal{D}_{-1/2}+\mathcal{D}_{1/2},
\end{equation}
with
\begin{equation}
    \mathcal{D}_l\simeq\pdv{}{\psi}\ln p+\frac{m(u+V_\parallel)}{T}\frac{\Omega}{I}+\left(\frac{mu^2}{2T}+l\right)\pdv{}{\psi}\ln T.
\end{equation}
Equation \eqref{D1} gives the equation for the parallel flow similar to \eqref{Gammawg zero}
\begin{equation}\label{Istvan Gamma zero}
\pdv{}{\psi}\ln p+\frac{m(u+V_\parallel)}{T}\frac{\Omega}{I}=-\frac{J_{PC}(U^2)}{2}\pdv{}{\psi}\ln T,
\end{equation}
where we have adopted the notation in \cite{Pusztai2010},
\begin{equation} \label{JPC}
    J_{PC}(U^2)=\frac{4U^6-2U^4+1}{1+2(U^2+U^4)},
\end{equation}
and $U^2=mu^2/2T$. Equation \eqref{JPC} is equivalent to (31) in \cite{Pusztai2010}. Similarly, the energy flux in this limit becomes
\begin{multline}
    Q=-\frac{3nTI^2}{\Omega^2 qR}\left(\frac{r}{R}\right)^2\sqrt{\frac{\pi}{2}}\left(\frac{T}{m}\right)^{3/2}\exp\left(-\frac{mu^2}{2T}\right)\Bigg\lbrace\frac{m^3u^6}{12T^3}\mathcal{D}_{-3/2}\\
        +\frac{m^2u^4}{3T^2}\mathcal{D}_{-1/2}        +\frac{5}{6}\frac{mu^2}{T}\mathcal{D}_{1/2}
        +\mathcal{D}_{3/2}\Bigg\rbrace,
\end{multline}
which can be written as
\begin{equation}
    Q=-\frac{3\sqrt{\pi}}{4}\frac{nT}{qR}\frac{I^2}{\Omega^2 }\left(\frac{r}{R}\right)^2
       \left(\frac{2T}{m}\right)^{3/2}\exp\left(-\frac{mu^2}{2T}\right)L_{PC}(U^2)\pdv{}{\psi}\ln T,
\end{equation}
where
\begin{equation}
    L_{PC}(U^2)=\frac{4U^8/3+16U^6/3+8U^4+4U^2+1}{1+2(U^2+U^4)},
\end{equation}
which agrees with (29) in \cite{Pusztai2010}. Lastly, we can compare the bootstrap current in this limit. We find
\begin{multline}
    \langle j_\parallel^B\rangle_\psi= -\frac{1}{8}\sqrt{\frac{\pi}{2}}\frac{en_e}{\nu_{ee}qR}\frac{I}{\Omega_e}\left(\frac{r}{R}\right)^2\left(\frac{2T_e}{m_e}\right)^{3/2}\bigg[2a_0\mathcal{D}_{e,1/2}-a_1\mathcal{D}_{e,13/2}\\
    -\sum_{i=2} \frac{a_i\Gammaf(i-\frac{1}{2})}{\sqrt{\pi}\Gammaf(1+i)}  \mathcal{D}_{e,\gamma_i} \bigg].
\end{multline}
We note that to compare with \cite{Pusztai2010}, their definition of $a_0$ and $a_1$ is different from ours by a factor $-\sqrt{2}$, so
\begin{align}
    a_0=-\sqrt{2}a_0^{PC}&&a_1=-\sqrt{2}a_1^{PC},
\end{align}
as their collision frequency is following the standard Braginskii form, thus $\nu_e^{PC}=\sqrt{2}\nu_{ee}$. Furthermore, \cite{Pusztai2010} only kept the first two elements of the Spitzer-Härm function. 
We need to substitute expression \eqref{Istvan Gamma zero} for the mean parallel flow to write the bootstrap current as
\begin{multline}
    \langle j_\parallel^B\rangle_\psi= -\frac{1}{8}\sqrt{\frac{\pi}{2}}\frac{en_e}{\nu_{ee}qR}\frac{I}{\Omega_e}\left(\frac{r}{R}\right)^2\left(\frac{2T_e}{m_e}\right)^{3/2}(2a_0-a_1)\bigg[\frac{1}{p_e}\pdv{p}{\psi}+\frac{2a_0-13a_1}{4a_0-2a_1}\pdv{}{\psi}\ln T_e\\
    +\frac{J_{PC}(U^2)}{2ZT_e}\pdv{T}{\psi}   \bigg],
\end{multline}
where we only kept the first two terms in the Spitzer-Härm function. This is in agreement with equation (50) in \cite{Pusztai2010}. \par
We note that our results are only consistent with those in \cite{Pusztai2010} if the comparison is made with their originally published results and not with the corrected results in the corrigendum. The reason for this is in the replacement of the collision operator with the Krook operator. In their original approach, the drift kinetic equation was written in their equation (12) as
\begin{equation}\label{Peter12}
    (Sv_\parallel+u_\ast)\bm{\hat{b}}\bcdot\bnabla\left[H_i+\frac{Iv_\parallel f_{M}}{\Omega}\left(\frac{mv^2}{2T}-\frac{5}{2}\right)\pdv{}{\psi}\ln T\right]-C^{(l)}[H_i]=0,
\end{equation}
where $H_i$ is defined in (13) as
\begin{equation}\label{Hi1}
    H_i=h_{1i}-\frac{Iv_\parallel f_M}{\Omega}\left(\frac{mv^2}{2T}-\frac{5}{2}\right)\pdv{}{\psi}\ln T.
\end{equation}
In the original work, the derivative of $v^2/2$ in \eqref{Peter12} was mistakenly assumed constant, although the total energy $E=v^2/2+Ze\Phi/m$ is held fixed. The derivative of $Ze\Phi/m$ was added in the corrigendum. In the next step, a term is added to $H_i$, $H_i\rightarrow H_i+mBkv_\parallel f_M/T$, to ensure that the flow is divergence-free before the collision operator is replaced with the Krook operator.\par 
Instead of following the procedure of \cite{Pusztai2010} one could imagine writing \eqref{Peter12} as
\begin{equation}\label{Peter12new}
    (Sv_\parallel+u_\ast)\bm{\hat{b}}\bcdot\bnabla\left[H_i+\frac{Iv_\parallel f_{M}}{\Omega}\left(\frac{mE}{T}-\frac{5}{2}\right)\pdv{}{\psi}\ln T\right]-C^{(l)}[H_i]=0,
\end{equation}
where now $H_i$ is defined as
\begin{equation} \label{Peter13new}
    H_i=h_{1i}-\frac{Iv_\parallel f_M}{\Omega}\left(\frac{mE}{T}-\frac{5}{2}\right)\pdv{}{\psi}\ln T.
\end{equation}
This is possible because at this point the collision operator still preserves momentum and $C[f_Mv_\parallel]=0$. One could now add a term to require zero divergence of flow and replace the collision operator with a Krook operator. The difference would be that now the derivative term of $E$ would not give a contribution as $E$ is held constant. If one chooses the second approach, one recovers the original form of $J_{PC}(U^2)$ that our results agree with. \par 
The choice of keeping or not keeping the difference between \eqref{Hi1} and \eqref{Peter13new}, appears to give different results. We argue that because the Krook operator does not preserve momentum, the replacement cannot be easily made, as adding a term proportional to $v_\parallel f_M$ inside the collision operator should not change the final result. It is possible that the additional term $mBkv_\parallel f_M/T$ does only ensure a divergence free flow if $H_i$ is of the form \eqref{Peter13new} which would effectively lead to the original results in \cite{Pusztai2010}. In our treatment, we do not use a simplified collision operator and thus we do not run the risk of breaking momentum conservation.

\subsection{Comparison with the work of \cite{Seol2012}}
In the work by \cite{Seol2012} on corrections to neoclassical theory in the plateau regime for strong gradients, strong gradients in density, temperature and electric field were considered. However, the gradient of the mean parallel flow was assumed to be small, and the poloidal variation of the potential was set to zero. Discrepancies between the works of \cite{Kagan2008, Pusztai2010, Catto2011} and \cite{Trinczek2023, trinczek2025a} and the works of \cite{Shaing1992, Shaing1994, Shaing2012, Seol2012} have been pointed out repeatedly across different collisionality regimes and also affect the comparison between this work and \cite{Seol2012}. The difference stems from a term proportional to $v^2-3v_\parallel^2$ that appears in the drift kinetic formulation of \cite{Seol2012}. In our work, the corresponding term is proportional to $v_\parallel^2+v^2$. We now show how this discrepancy arises due to the strong gradient expansion in the moment approach to neoclassical theory which is used in \cite{Shaing1992, Shaing1994, Shaing2012, Seol2012}.\par
The moment approach to neoclassical theory uses a decomposition of the distribution function $f=f_M+f_1$ into
\begin{equation}
    f_1=-\frac{Iv_\parallel}{\Omega}\pdv{f_M}{\psi}+f_M\frac{mv_\parallel}{T}\sum_{j=0}^\infty a_jL_j^{(3/2)}(x^2).
\end{equation}
Here, the Maxwellian is a function of energy. The first two coefficients in the expansion for $f_1$ are $a_0=V^\theta B$ and $a_1=-2q^\theta B/5p$, where the quantities $V^\theta$ and $q^\theta$ are only functions of $\psi$,
\begin{align}\label{Vpar and qpar}
    V_\parallel=V_1+V^\theta(\psi)B,&&q_\parallel=\frac{5p}{2}V_2+q^\theta(\psi)B
\end{align}
with
\begin{align}
    V_1\equiv-\frac{IT}{m\Omega}\left(\pdv{}{\psi}\ln p+\frac{Ze}{T}\pdv{\Phi}{\psi}\right),&&V_2\equiv-\frac{IT}{m\Omega}\pdv{}{\psi}\ln T.
\end{align}
One can show that $V_\parallel$ and the parallel heat flux $q_\parallel$ have this form by imposing vanishing divergence of flows \citep{helander2005}. 
It is useful in the moment approach to write the expansion of $f_1$ in the form
\begin{equation}\label{f1 moment approach}
    f_1= f_M\frac{mv_\parallel}{T}V^\theta(\psi)B-f_M\frac{mv_\parallel}{T}\frac{2}{5p}q^\theta(\psi)B L_1^{(3/2)}(x^2)-\frac{Iv_\parallel}{\Omega}\pdv{f_M}{\psi}+h,
\end{equation}
where 
\begin{equation}
    h=f_M\frac{mv_\parallel}{T}\sum_{j=2}^\infty a_jL_j^{(3/2)}(x^2).
\end{equation}
All $\theta$-dependence in \eqref{f1 moment approach} is in the factors $v_\parallel B$ and $v_\parallel/B$, and in $h$. The drift kinetic equation that we need to solve for $f_1$ is
\begin{equation}\label{drift eq moment approach}
    v_\parallel\bm{\hat{b}}\bcdot\nabla f_1+\bm{v_d}\bcdot\nabla f_1+\bm{v_d}\bcdot\nabla \psi \pdv{f_M}{\psi}=C^{(l)}[f_1].
\end{equation}
We can use the four identities
\begin{equation}\label{Iden1}
    v_\parallel \bm{\hat{b}}\bcdot\bnabla(v_\parallel B)=v^2 \left(\frac{3 v_\parallel^2}{2v^2}-\frac{1}{2}\right)\bm{\hat{b}}\bcdot\nabla B
\end{equation}

\begin{equation}
    v_\parallel \bm{\hat{b}}\bcdot\bnabla\left(\frac{v_\parallel}{ B}\right)=-v^2 \left(\frac{v_\parallel^2}{2v^2}+\frac{1}{2}\right)\frac{\bm{\hat{b}}\bcdot\nabla B}{B^2}
\end{equation}
\begin{equation}
    \bm{v_d}\bcdot\bnabla\psi=-\frac{I}{B\Omega}\left(\frac{v^2}{2}+\frac{v_\parallel^2}{2}\right)\bm{\hat{b}}\bcdot\nabla B
\end{equation}
and 
\begin{equation}\label{Iden4}
    \bm{v_d}\bcdot\bnabla\theta\simeq u\bm{\hat{b}}\bcdot\nabla\theta.
\end{equation}
to combine the expression for $f_1$ \eqref{f1 moment approach} and the drift kinetic equation \eqref{drift eq moment approach} and find

\begin{multline}\label{D25}
    \left(v_\parallel+u\right)\bm{\hat{b}}\bcdot\bnabla\theta \pdv{h}{\theta} +\bm{v_d}\bcdot\bnabla\psi\pdv{h}{\psi}\\
    +\left(1+\frac{u}{v_\parallel}\right)\left(\frac{3 v_\parallel^2}{2}-\frac{v^2}{2}\right)\bm{\hat{b}}\bcdot\bnabla B\left[ \frac{2}{v_{t}^2}\left(V^\theta-\frac{2}{5p}L_1^{(3/2)}(x^2)q^\theta\right)f_M \right]\\
    -\left(1+\frac{u}{v_\parallel}\right)\left(\frac{v_\parallel^2}{2}+\frac{v^2}{2}\right)\frac{\bm{\hat{b}}\bcdot\bnabla B}{B}\left[\frac{I}{\Omega}\left(-\pdv{}{\psi}\ln p-\frac{Ze}{T}\pdv{\Phi}{\psi}+L_1^{(3/2)}(x^2)\pdv{}{\psi}\ln T \right)f_M\right]\\
    -\frac{I}{B\Omega}\left(\frac{v_\parallel^2}{2}+\frac{v^2}{2}\right)\bm{\hat{b}}\bcdot\bnabla B \left(\pdv{}{\psi}\ln p+\frac{Ze}{T}\pdv{\Phi}{\psi}-L_1^{(3/2)}(x^2)\pdv{}{\psi}\ln T \right)f_M\\
    +\bm{v_d}\bcdot\bnabla\psi \pdv{}{\psi}\left[ \frac{2v_\parallel}{v_{t}^2}\left(V_\parallel-\frac{2}{5p}L_1^{(3/2)}(x^2)q_\parallel\right)f_M \right]=C^{(l)}[h].
\end{multline}
One can neglect the last term on the left hand side if $V_\parallel$ and $q_\parallel$ are assumed to be small. We note that the terms in the third and fourth line of \eqref{D25}, both proportional to $v_\parallel^2+v^2$, almost cancel with each other for small $u$,
\begin{multline}\label{drift f1 momentum}
\left(v_\parallel+u\right)\bm{\hat{b}}\bcdot\bnabla\theta \pdv{h}{\theta} +\bm{v_d}\bcdot\bnabla\psi\pdv{h}{\psi}\\
+\left(1+\frac{u}{v_\parallel}\right)\left(\frac{3 v_\parallel^2}{2}-\frac{v^2}{2}\right)\bm{\hat{b}}\bcdot\bnabla B\left[ \frac{2}{v_{t}^2}\left(V^\theta-\frac{2}{5p}L_1^{(3/2)}(x^2)q^\theta\right)f_M \right]\\
    -\frac{u}{v_\parallel}\left(\frac{v_\parallel^2}{2}+\frac{v^2}{2}\right)\frac{\bm{\hat{b}}\bcdot\bnabla B}{B}\left[\frac{I}{\Omega}\left(-\pdv{}{\psi}\ln p-\frac{Ze}{T}\pdv{\Phi}{\psi}+L_1^{(3/2)}(x^2)\pdv{}{\psi}\ln T \right)f_M\right]=C^{(l)}[h].
\end{multline}
Indeed, if we take the limit of weak gradients, i.e. $u\rightarrow 0$, we recover
\begin{equation}\label{wg limit}
    v_\parallel \bm{b}\bcdot\bnabla\theta \pdv{h}{\theta}-C[h]=\frac{2v^2}{v_{t}^2}\left(\frac{1}{2}-\frac{3 v_\parallel^2}{2v^2}\right)\bm{\hat{b}}\bcdot\bnabla B\left(V^\theta-\frac{2}{5p}L_1^{(3/2)}(x^2)q^\theta\right)f_M .
\end{equation}
However, if we take the limit of strong radial electric field, trapped particles satisfy $v_\parallel\simeq-u$, and we find
\begin{multline}\label{drift eq moment 2}
(v_\parallel+u)\bm{\hat{b}}\bcdot\bnabla\theta \pdv{h}{\theta}+\bm{v_d}\bcdot\bnabla\psi\pdv{h}{\psi}-C[h]\\
=-\left(\frac{v_\parallel^2}{2}+\frac{v^2}{2}\right)\frac{\bm{\hat{b}}\bcdot\bnabla B}{B}\left[\frac{I}{\Omega}\left(-\pdv{}{\psi}\ln p-\frac{Ze}{T}\pdv{\Phi}{\psi}+L_1^{(3/2)}(x^2)\pdv{}{\psi}\ln T \right)f_M\right].
\end{multline}
If $V_\parallel$ and $q_\parallel$ are assumed to be small, as in \cite{Seol2012} and \cite{Shaing2012}, then $V_1\simeq-V^\theta B$ and $5pV_2/2\simeq-q^\theta B$. We find that the drift kinetic equation for strong gradients in the trapped particle region with small flows assumed should read
\begin{multline}\label{drift eq moment 3}
(v_\parallel+u)\bm{\hat{b}}\bcdot\bnabla\theta \pdv{h}{\theta}+\bm{v_d}\bcdot\bnabla\psi\pdv{h}{\psi}-C[h]\\
=\frac{2v^2}{v_{t}^2}\left(\frac{v_\parallel^2}{2v^2}+\frac{1}{2}\right)\bm{\hat{b}}\bcdot\bnabla B\left(V^\theta-\frac{2}{5p}L_1^{(3/2)}(x^2)q^\theta\right)f_M .
\end{multline}
We see the difference between the source proportional to $v^2-3v_\parallel^2$ in the weak gradient limit \eqref{wg limit}, and the source proportional to $v_\parallel^2+v^2$ in the strong gradient limit \eqref{drift eq moment 3}.
In the works by \cite{Seol2012, Shaing2012}, the equation used to solve for $f_1$ is
\begin{multline}
    (v_\parallel+u)\bm{\hat{b}}\bcdot\bnabla \theta\pdv{h}{\theta}+\bm{v_d}\bcdot\bnabla\psi\pdv{h}{\psi}-C[h]\\
    =\frac{2v^2}{v_{t}^2}\left(\frac{1 }{2}-\frac{3 v_\parallel^2}{2 v^2}\right)\bm{\hat{b}}\bcdot\bnabla B\left(V^\theta-\frac{2}{5p}L_1^{(3/2)}(x^2)q^\theta\right)f_M .
\end{multline}
We thus conclude that the right-hand side of the drift kinetic equation is the one from the weak gradient limit and is missing the strong radial electric field terms in the trapped region that would have given rise to the terms proportional to $(v_\parallel^2+v^2)$. The results by \cite{Seol2012, Shaing2012} need to be corrected for this discrepancy. Furthermore, these authors did not retain poloidal variation or $V_\parallel\sim v_{t}$. Importantly, we have shown that the small flow assumption $V_\parallel\ll v_{t}$ is not consistent in strong gradient regions, as it would not be possible to satisfy \eqref{Maxwellian} without $V_\parallel\sim v_t$.

\section{Bootstrap current derivation}\label{sec: bootstrap Appendix}
The bootstrap current in the plateau regime follows from inserting the distribution function \eqref{ge0} into \eqref{bootstrap integral}. The integral over $v_\parallel$ is identical to the integral \eqref{int gl} with coefficients $A_{Ce}$ and $A_{Se}$ instead of $A_C$ and $A_S$. We are left with the integral
\begin{equation}\label{bootstrap general explicit}
    \langle j_\parallel^B\rangle_\psi^{wg}=- \frac{e}{4\nu_{ee}} \frac{\sqrt{\pi} }{ qR}\left(\frac{r}{R}\right)^2\frac{I}{\Omega}n_e\sqrt{\frac{m_e}{ T_e}}\sum_{i=0}a_i\int\mathrm{d}(x_e^2)\:L^{3/2}_i\left(x_e^2\right)e^{-x_e^2}\left(A_{Se}^2+A_{Ce}^2 \right)\mathcal{D}_e ,
\end{equation}
In order to integrate \eqref{bootstrap general explicit}, we need to use that
\begin{equation}
    \int_0^\infty\mathrm{d}x\: L_i^{3/2}(x) e^{-x} =\frac{2\Gammaf(i+\frac{3}{2})}{\sqrt{\pi}\Gamma(1+i)},
\end{equation}
\begin{equation}
    \int_0^\infty\mathrm{d}x\: L_i^{3/2}(x) e^{-x} L_1^{3/2}(x) =\frac{\left(\frac{3}{2}+5i\right)\Gammaf(i+\frac{1}{2})}{\sqrt{\pi}\Gammaf(1+i)},
\end{equation}

\begin{equation}
    \int_0^\infty\mathrm{d}x\: L_i^{3/2}(x) e^{-x} L_1^{3/2}(x)x =\frac{\left(-1+10i\right)\Gammaf(i-\frac{1}{2})}{4\sqrt{\pi}\Gammaf(1+i)}\quad\text{for } i\geq1,
\end{equation}
and
\begin{equation}
    \int_0^\infty\mathrm{d}x\: L_0^{3/2}(x) e^{-x} L_1^{3/2}(x)x =\frac{1}{2},
\end{equation}
as well as 
\begin{equation}
    \int_0^\infty\mathrm{d}x\: L_i^{3/2}(x) e^{-x} L_1^{3/2}(x)x^2 =-\frac{\left(3+10i\right)\Gammaf(i-\frac{3}{2})}{4\sqrt{\pi}\Gammaf(1+i)}\quad\text{for } i\geq2,
\end{equation}
and
\begin{align}
    \int_0^\infty\mathrm{d}x\: L_1^{3/2}(x) e^{-x} L_1^{3/2}(x)x^2 =\frac{13}{2},&&\int_0^\infty\mathrm{d}x\: L_0^{3/2}(x) e^{-x} L_1^{3/2}(x)x^2 =-1.
\end{align}
We can use $x=5/2-L_1^{3/2}(x)$, $x^2=2L_2^{3/2}(x)-7L_1^{3/2}(x)+35/4$ and the identities above to find
\begin{equation}
      \int_0^\infty\mathrm{d}x\: L_i^{3/2}(x) e^{-x}x=\frac{\Gammaf(i+\frac{1}{2})}{\sqrt{\pi}\Gammaf(i+1)}.
\end{equation}
and
\begin{equation}
      \int_0^\infty\mathrm{d}x\: L_i^{3/2}(x) e^{-x}x^2=-\frac{\Gammaf(i-\frac{1}{2})}{\sqrt{\pi}\Gammaf(i+1)} \quad \text{for } i\geq 1
\end{equation}
and
\begin{equation}
      \int_0^\infty\mathrm{d}x\: L_0^{3/2}(x) e^{-x}x^2=2.
\end{equation}
We can substitute these identities into \eqref{bootstrap general explicit} to find the result for the bootstrap current \eqref{bootstrap}.

\section{Normalised amplitudes of the poloidal variation}\label{sec: Transport equations}

The quasineutrality relations \eqref{phic plateau} and \eqref{phis plateau} yield the amplitudes of the poloidal variation of the electric potential
\begin{align}\label{HolyPhi}
    \bar{\phi}_c=\frac{EA+GD}{A^2+Z\bar{T}_eD^2} && \text{and}&&\bar{\phi}_s=\frac{GA-Z\bar{T}_e ED}{A^2+Z\bar{T}_eD^2},
\end{align}
where
\begin{multline}
    A= 1-\frac{Z\bar{T}_e}{\bar{T}}\bigg[\sqrt{\bar{T}}J\left(\pdv{}{\bar{\psi}}\ln\bar{p}-\frac{3}{2}\pdv{}{\bar{\psi}}\ln\bar{T}\right)\\
    +\left(1-2\frac{\bar{V}+\bar{u}}{\sqrt{\bar{T}}}J\right)\left(\pdv{\bar{V}}{\bar{\psi}}-1-\frac{\bar{V}+\bar{u}}{2}\pdv{}{\bar{\psi}}\ln\bar{T}\right)\bigg],
\end{multline}
\begin{multline}
    E=-Z\bar{T}_e \Bigg\lbrace\sqrt{\bar{T}}J\left[\left(\frac{2\bar{V}^2}{\bar{T}}+1\right)\left(\pdv{}{\bar{\psi}}\ln\bar{p}-\frac{3}{2}\pdv{}{\bar{\psi}}\ln\bar{T}\right)+\pdv{}{\bar{\psi}}\ln \bar{T}\right]+\left(1-2\frac{\bar{V}+\bar{u}}{\sqrt{\bar{T}}}J\right)\\
    \times\bigg[(\bar{V}-\bar{u})\left(\pdv{}{\bar{\psi}}\ln\bar{p}-\frac{3}{2}\pdv{}{\bar{\psi}}\ln \bar{T}\right)+\left(\pdv{\bar{V}}{\bar{\psi}}-1\right)\left(\frac{2\bar{u}^2}{\bar{T}}+1-\frac{2(\bar{V}+\bar{u})^2}{\bar{T}}\right)\\
    -\frac{\bar{V}+\bar{u}}{2}\pdv{}{\bar{\psi}}\ln\bar{T}\left(\frac{2\bar{V}}{\bar{T}}+1\right)\bigg]\\
    +\left[1+2\frac{(\bar{V}+\bar{u})^2}{\bar{T}}-4\frac{(\bar{V}+\bar{u})^{3}}{\bar{T}^{3/2}}J\right]\left(\pdv{\bar{V}}{\bar{\psi}}-1+\frac{\bar{V}-\bar{u}}{2}\pdv{}{\bar{\psi}}\ln \bar{T}\right)+2\bigg\rbrace,
\end{multline}
\begin{equation}
    D=-\sqrt{\frac{\pi}{\bar{T}}}\exp\left[-\frac{(\bar{u}+\bar{V})^2}{\bar{T}}\right]\bar{\mathcal{D}}_{-3/2},
\end{equation}
and
\begin{equation}
    G=Z\bar{T}_e\sqrt{\frac{\pi}{\bar{T}}}\exp\left[-\frac{(\bar{u}+\bar{V})^2}{\bar{T}}\right]\left(\bar{u}^2\bar{\mathcal{D}}_{-3/2}+\frac{\bar{T}}{2}\bar{\mathcal{D}}_{-1/2}\right).
\end{equation}
\par

\bibliographystyle{jpp}

\bibliography{MyLibrary.bib}
\end{document}